\documentclass[preprint,showpacs,preprintnumbers,amsmath,amssymb]{revtex4}

\usepackage{graphicx}
\usepackage{dcolumn}
\usepackage{bm}
\usepackage[final]{pdfpages}

\newcommand{\bq}{\begin{equation}}
\newcommand{\eq}{\end{equation}}
\newcommand{\bqa}{\begin{eqnarray}}
\newcommand{\eqa}{\end{eqnarray}}
\newcommand{\nn}{\nonumber \\}

\def\be     {\begin{equation}}
\def\ee     {\end{equation}}
\def\bea        {\begin{eqnarray}}
\def\eea        {\end{eqnarray}}
\def\bnn    {\begin{eqnarray*}}
\def\enn    {\end{eqnarray*}}

\begin{document}



Introduction

It is rare to observe Weyl fermions in three dimensional condensed matter systems. Graphene may be one of the well-studied systems as a two-dimensional Weyl metallic state, where two types of Weyl fermions with opposite chirality appear at separate momenta [1]. However, no experimental evidence has been reported for the Weyl metallic phase in three dimensions, where they are combined more often to form Dirac fermions. Rigorously speaking, this phenomenon is attributed to the fact that the Dirac spinor is the irreducible representation of the Lorentz group in three dimensions while it is the Weyl spinor in two dimensions [2]. Recently, it has been shown that such a Weyl metallic phase must exist near the topological phase transition from a band insulator to a topological insulator in three dimensions when either time reversal symmetry or inversion symmetry is not preserved at the critical point [3-7]. Actually, the emergence of Weyl fermions has been claimed long before when magnetic fields are applied to gapless semi-conductors (Dirac metals) [9,10]. A Dirac point is separated into two Weyl points with opposite chirality in momentum space, where their distance is proportional to the applied magnetic field.

The characteristic feature of a Weyl metallic phase is that a Weyl fermion state at one Weyl point must transfer to that at the other Weyl point when currents are driven in the same direction as the momentum to connect two paired Weyl points. This phenomenon is referred to as the Adler-Bell-Jackiw anomaly, which means that the chiral current is not conserved [9,10]. This nonlocal (in momentum space), more accurately, topological "constraint" has been suggested to cause anomalous magneto-transport phenomena in three dimensional Weyl metals, that is, "negative" magnetoresistivity (MR) when the applied electric field is parallel to the magnetic field. This anomalous longitudinal MR was shown to arise from the suppression of scattering between two Weyl points as the magnetic fields are enhanced [8,11]. Unfortunately, this Weyl metallic phase has not been confirmed experimentally. In particular, the "negative" MR (or "positive" magnetoconductivity-MC) has not been observed yet.

In this letter we claim that Weyl fermions appear in $Bi_{1-x}Sb_x$ near the critical point of the topological phase transition when magnetic fields are applied. The global phase diagram of $Bi_{1-x}Sb_x$ is well known [12-14] (Fig. 1(a)). A topological phase transition from a band insulator to a topological insulator occurs at $x \approx 3\%$, where massless Dirac fermions appear near the L point. Therefore, $Bi_{1-x}Sb_x$ at $x \approx 3\%$ is identified as a Dirac metal. Applying magnetic fields near this critical point, it is natural to expect the emergence of a Weyl metal phase, where the single Dirac point splits into two Weyl points with opposite chirality. We verify the emergence of the Weyl metallic phase by measuring the angle dependence of MR, where the angle corresponds to that between applied magnetic and electric fields. In particular, we observe not only the weak anti-localization near zero

magnetic fields ($B < 0.4$ T) but also an upturn of or increasing MC above 0.4 T only when both magnetic and electric fields are applied in the same direction. The coexistence between weak anti-localization and "negative" MR is remarkable. The experimental observation of weak anti-localization corrections is common for systems with strong spin-orbit interactions [15-19]. On the other hand, negative MR results from either weak localization or interaction corrections with diffusive dynamics (diffusive Fermi liquids) [20]. In addition, it can appear when magnetic fluctuations are suppressed via magnetic fields, and thus the corresponding scattering rate decreases [21]. Considering the fact that conventional calculations based on density functional theory describe the band structure of $Bi_{1-x}Sb_x$ quite well [12,13], such interaction effects cannot be the physical origin for the "negative" MR. Then, their coexistence is difficult to understand within the perturbation framework for both nonmagnetic randomness and weak interactions in the presence of spin-orbit scattering.

We attribute the underlying mechanism for the upturn of MC to the Adler-Bell-Jackiw anomaly in the presence of weak anti-localization corrections, comparing experimental data with theoretical results from the quantum Boltzmann equation approach with the introduction of the topological $E \cdot B$ term [22]. In the "longitudinal" magneto-transport configuration where the electric field is parallel to the magnetic field, the dynamics of Weyl fermions is topologically constrained by the topological $E \cdot B$ term [8,11]. As a result, the MC increases above a critical magnetic field associated with the weak anti-localization. In contrast, only weak anti-localization corrections to the MC were found in the conventional "transverse" magneto-transport measurements, where the magnetic and electric fields were applied in the $z$ and $x$ directions, respectively. This paper suggests the unexpected coexistence of weak anti-localization and the upturn of "longitudinal" MC as the fingerprints of Weyl fermions in three dimensions, which arise from the topological $E \cdot B$ term or the Adler-Bell-Jackiw anomaly.

Experiments and analysis

Figure 1(b) presents a schematic diagram of the experiments for measuring the electrical transport coefficients, $\rho_{\alpha\beta}(B_\gamma)$, which are defined by the equation, $E_\alpha = \rho_{\alpha\beta}(B_\gamma)J_\beta$, where $J_\beta$ is an electric current of the $\beta$ direction under a magnetic field $B_\gamma$ in the $\gamma$ direction and $E_\alpha$ is the induced electric field in the $\alpha$ direction. In these experiments, six different field-contact-current configurations were employed, where the currents were set to flow along the $x$ direction, voltage contacts were located either on the $x$ (MR configuration) or $y$ axis (Hall configuration), and magnetic fields were applied along all three directions. $\rho_{xx}(B_z)$ and $\rho_{yx}(B_z)$, which are called transverse MR and transverse Hall resistivity, respectively, are regarded as a conventional setup, where the currents and

magnetic fields are orthogonal to each other in the out-of-plane. $\rho_{xx}(B_x)$ and $\rho_{yx}(B_x)$ are longitudinal MR and longitudinal Hall resistivity, respectively, where the currents and magnetic fields are parallel to each other. $\rho_{xx}(B_y)$ and $\rho_{yx}(B_y)$ are unconventional transverse MR and unconventional transverse Hall resistivity, respectively, where the currents and magnetic fields are orthogonal to each other but in plane. The unconventional transverse MR of $\rho_{xx}(B_y)$ in Fig. 1(b)-(5) gives rise to similar results to the transverse MR of $\rho_{xx}(B_z)$ in Fig. 1(b)-(1), provided the band anisotropy is not too large. In this study we focus on $\rho_{xx}(B_z)$ in Fig. 1(b)-(1) and $\rho_{xx}(B_x)$ in Fig. 1(b)-(3) and leave the role of the $E \cdot B$ term in $\rho_{yx}(B_x)$ in Fig. 1(b)-(4) and $\rho_{yx}(B_y)$ in Fig. 1(b)-(6) as an important future work.

Figure 2 shows the angle-dependent MR for the configuration, as shown in the schematic diagram, where the configurations for θ=0° and 90° were used to measure the transverse [Fig. 1(b)-(1)] and longitudinal [Fig. 1(b)-(3)] MRs, respectively. All the MR data showed a narrow dip below 0.4 T. These dips are similar to those observed in graphene [16,17] and thin films of three-dimensional topological insulators [18,19], which were attributed to weak anti-localization. On the other hand, an important difference was observed between the previous studies and the present measurements; the weak anti-localization in the data is due to three-dimensional Dirac fermions instead of two-dimensional ones. It is noted that $x \approx 3\%$ corresponds to the critical point, where three-dimensional Dirac fermions appear. One way to confirm the three-dimensionality is to determine if the scaling property is satisfied for different angles. If the dynamics of Dirac fermions is two dimensional, only the magnetic field perpendicular to the plane for the dynamics of such Dirac fermions leads to cyclotron motions. This scaling behavior was not observed in $Bi_{0.97}Sb_{0.03}$, supporting the three-dimensional nature of Dirac fermions (Fig. 2(b)). The MR data will be fitted to the three dimensional expression for weak anti-localization later.

Unexpected features were noted at higher magnetic fields, particularly for the longitudinal MR configuration of θ = 90°. When θ is near 0°, i.e. a conventional setup for transverse MR, the MR increased quadratically, indicating the dominance of orbital contributions. This standard behavior changed drastically when θ approached 90°, where the MR decreases considerably just after the dip up to $B \sim 4$ T. Above 4 T, MR increases again. As discussed before, this "negative" MR is difficult to understand within the perturbation framework incorporating both electronic (magnetic) correlations and nonmagnetic impurities in the presence of spin-orbit interactions. The decrease in longitudinal MR at relatively moderate magnetic fields above the region of weak anti-localization is attributed to the topologically constrained dynamics of Weyl fermions, which is given by the chiral anomaly.

Both $\rho_{xx}(B_z)$ and $\rho_{xx}(B_x)$ are derived based on the quantum Boltzmann equation approach with the introduction of the Adler-Bell-Jackiw anomaly via the semi-classical equations of motion [22,23]. Unfortunately, technical complexity of this methodology does not allow us to have an intuitive physical picture for anomalous behaviors in such transport coefficients. To make physics more apparent, we re-derive these anomalous transport coefficients based on the semi-classical equations of motion only. This approach is essentially the same as the equation of motion with the Lorentz force for the Hall effect in the elementary solid-state physics except for the introduction of the topological $E \cdot B$ term and the contribution of the Berry curvature. All details are discussed in the supplementary material [23] and a recent paper [22].

Disappearance of Weyl fermions with – chirality gives rise to the production of those with +, where this dissipationless transfer is equilibrated by several scattering processes to reach a steady state. The Adler-Bell-Jackiw anomaly and possible scattering channels are described by the semi-classical equation of motion for the momentum, $\dot{p} = \left(1 + \frac{e}{c} B \cdot \Omega_p\right)^{-1} \left\{ eE + \frac{e}{mc} p \times B + \frac{e^2}{c}(E \cdot B)\Omega_p \right\} = -\frac{p}{\tau}$, where $\tau$ is an effective mean-free time, determined by the intra-node and inter-node scattering times (node = Weyl point) [23]. Solving this equation of motion to obtain the momentum as a function of both electric and magnetic fields, we find the corresponding electrical current, $J = ne\dot{r} = ne\left(1 + \frac{e}{c} B \cdot \Omega_p\right)^{-1} \left\{ \frac{p}{m} + eE \times \Omega_p + \frac{e}{mc}(\Omega_p \cdot p)B \right\}$, where the solution of the velocity was utilized from the other semi-classical equation of motion [23]. It is essential to notice that there are two additional contributions for electric currents other than the conventional term proportional to the momentum, nonzero only when the Berry curvature exists [24, 25].

After several straightforward steps with the condition for the Hall effect, $J_y = 0$, we find the expression of $\sigma_L(B) = (1 + C_W B^2) \cdot \sigma_{WAL} + \sigma_n$ for the longitudinal MC in the weak field region, essentially the same as that from the quantum Boltzmann equation approach, where $\sigma_{WAL}$ is the conductivity from weak anti-localization corrections associated with intra-node scattering and $\sigma_n$ is that from conventional Fermi surface contributions other than the L-point "Dirac" cone [23]. The most important feature is that the longitudinal MC contains an overall factor of $C_W B^2$ with a positive constant $C_W$, originating from the topological $E \cdot B$ term. Such a topological term turns out to cause an additional contribution for the momentum of the z-direction, driven by the x-directional electric field along with the x-directional magnetic field under the influence of Berry curvature [23]. Combined with the last term in the expression of the current, this contribution gives rise to the overall factor proportional to $B^2$, which enhances the longitudinal conductivity due to the momentum transfer

proportional to the applied magnetic field. In contrast, the transverse MC is expressed by $\sigma_T(B) = \sigma_{WAL} + \sigma_n$ without the anomaly contribution because the contribution from $E \cdot B$ vanishes. $\sigma_{WAL}$ and $\sigma_n$ are additive because they originate from different bands [26]. Considering that $\sigma_n$ would be determined by residual charge carriers around the T point in momentum space, the conventional Fermi-liquid form, $\sigma_n^{-1} = \rho_0 + A \cdot B^2$, is assumed, where the residual resistivity and the coefficient $A$ are determined from the best fit. Recalling the scaling result for the angle dependence of MR, the three-dimensional weak anti-localization formula, given by $\sigma_{WAL} = \Delta\sigma_{WAL} + \sigma_0 = a\sqrt{B} + \sigma_0$, where $a$ and $\sigma_0$ are also determined from the line of best fit, is used [27].

Figures 3(a) and 3(b) show the transverse MC $\sigma_T(B)$ and longitudinal MC $\sigma_L(B)$, respectively, where the black circles represent the experimental data and the red lines show the theoretical fitting based on the above equations. The essential features of the transverse MC, such as the sharp peak in the zero field region and the gradual decrease up to 1.2 T are reproduced quite well with parameters of $a$ = -14.3 $\Omega^{-1} \cdot T^{-0.5}$, $\sigma_0$ = 49.6 $\Omega^{-1}$, $\rho_0$ = 4.15×10$^{-2}$ $\Omega$, and A = 21.8 $\Omega \cdot T^{-2}$. We restrict the region of our fitting with -1.2 T < $B$ < 1.2 T in order to compare this analysis with the case of the longitudinal MC, where this formula is derived in the weak field region. This successful fitting confirms that the origin of the sharp peak is the three dimensional weak anti-localization. Similarly, the peak in the zero field region and the upturn of the longitudinal MC are also reproduced well with a reasonable set of parameters of $a$ = -20.6 $\Omega^{-1} \cdot T^{-0.5}$, $\sigma_0$ = 36.6 $\Omega^{-1}$, $\rho_0$ = 2.65×10$^{-2}$ $\Omega$, A = 7.3×10$^{-3}$ $\Omega \cdot T^{-2}$, and $C_W$ = 1.27 T$^{-2}$. All parameters of $a$, $\sigma_0$ and $\rho_0$ have similar values for both the transverse and longitudinal MCs, whereas the value of A in the longitudinal MC is much smaller than that in the transverse MC. This is consistent with our expectation, where the coefficient of the $B^2$ term in the normal conductivity from the bands around the T point should vanish because there cannot be any orbital motion in the longitudinal MC setup. The upturn of the longitudinal MC above 0.4 T cannot be captured without the correction term of $C_W B^2$, suggesting that the origin of this enhancement is purely topological, i.e. the chiral anomaly in the dynamics of Weyl fermions.

The positive component of the longitudinal MC disappears eventually at the critical magnetic field of $B \sim 4$ T and the MC decreases with increasing $B$ above 4 T. This "reentrant" downturn of MC may be discussed based on the "pair-annihilation" scenario [23]. However, a cautious person may criticize that it is unrealistic to consider such paired Weyl points to move all over the whole Brillouin zone away from the L point, because the band gap is rather large for the wave vectors away from the L point. Although our experiments are in the semi-classical regime associated with the Adler-Bell-Jackiw anomaly [24], it is difficult to exclude the possibility for the formation of Landau levels,

increasing magnetic fields to reach the "intermediate" region of the reentrant downturn behavior. This problem on the crossover from the semi-classical regime to the quantum regime is beyond the scope of the present investigation, which needs to be examined more sincerely near future both experimentally and theoretically.

Finally, we remark two different regimes for the Adler-Bell-Jackiw anomaly. The present analysis of the semi-classical regime is justified when the chemical potential is much larger than the cyclotron frequency. On the other hand, the so called ultra-quantum limit is realized when both the chemical potential and temperature are less than the energy gap between the lowest and first Landau levels [8,11,28]. Since only chiral branches of the spectrum are occupied, dynamics of these electrons are essentially the same as that of one-dimensional chiral fermions, where intra-node scattering is prohibited [23]. As a result, the effect of the Adler-Bell-Jackiw anomaly becomes enhanced, where the longitudinal current can be relaxed by inter-node scattering only. In this ultra-quantum limit, the correction term of the longitudinal MC is linearly proportional to $B$ due to one-dimensional chiral dynamics, distinguished from the case of the semi-classical regime [23]. All analysis based on the ultra-quantum limit failed to explain our experimental data consistently, indicating that Fermi levels of our samples are located far away from Weyl points [23]. Actually, the location of the chemical potential is consistent with our previous studies on anomalous transport phenomena [29,30].

Conclusion

Either time reversal or inversion symmetry breaking in three dimensional Dirac metals gives rise to Weyl metals, where the Dirac point described by the four-component Dirac spinor splits into two Weyl points described by the two-component Weyl spinors with opposite chirality. The dynamics of these Weyl fermions is constrained topologically when their currents are applied in the same direction as the momentum to connect the two Weyl points, referred to as the Adler-Bell-Jackiw anomaly and described by the topological $E \cdot B$ term. In the present study, the emergence of such Weyl fermions was confirmed from anomalous transport phenomena, applying magnetic fields near the topological phase transition from a band insulator to a topological insulator in $Bi_{1-x}Sb_x$. In particular, the upturn behavior above 0.4 $T$ in magnetoconductivity was observed ***only*** for $B \mathbin{/\mkern-5mu/} E$ besides weak anti-localization phenomena near zero magnetic fields ($B < 0.4\ T$). Both the quantum Boltzmann equation approach and the equation of motion approach with the introduction of the topological $E \cdot B$ term were used to derive a general expression for the longitudinal magnetoconductivity in the presence of weak anti-localization corrections, which reproduced our experimental data consistently.

Figure captions

Fig. 1 **Schematic band structure of $Bi_{1-x}Sb_x$ near the L point and schematic diagram of the experiments in this study.** (a) A topological phase transition occurs from a band insulator ($x < 3\%$) to a topological insular ($x > 3\%$) around $x \approx 3\%$, where band touching emerges to form a Dirac cone at the L point. Applying magnetic fields to this Dirac metal, the Dirac point described by the four-component Dirac spinor splits into two Weyl points described by the two-component Weyl spinors with opposite chirality. The distance between these two Weyl points is proportional to the applied magnetic field. (b) Schematic diagram of the electrical transport measurements. See the text for detailed explanation.

Fig. 2 **Angle-dependent (θ) magnetoresistance (MR) and its scaling property.** (a) Angle-dependent MR. The experimental configuration is depicted in (c). See the text for detailed explanation. (b) The scaling property of MR in the low-field region. The *x*-axis is the perpendicular field component of $B \cos θ$. This *x*-axis scaling fails to map the MR data of all θs into a single curve, indicating the three-dimensional nature of the low-field dip in MR. This dip was attributed to the three dimensional weak anti-localization.

Fig. 3 **Transverse and longitudinal magnetoconductances with theoretical fitting.** (a) Transverse magnetoconductance. The black circles represent the experimental data and the red lines express the theoretical fitting, where both weak anti-localization corrections and normal contributions from other bands are considered. See the text for detailed explanation. (b) Longitudinal magnetoconductance. The overall factor $C_W B^2$ in front of weak anti-localization corrections is the key feature, which originates solely from the topological $E \cdot B$ term in both the quantum Boltzmann equation and the equation of motion approaches. This is the fingerprint of the Weyl metallic state.



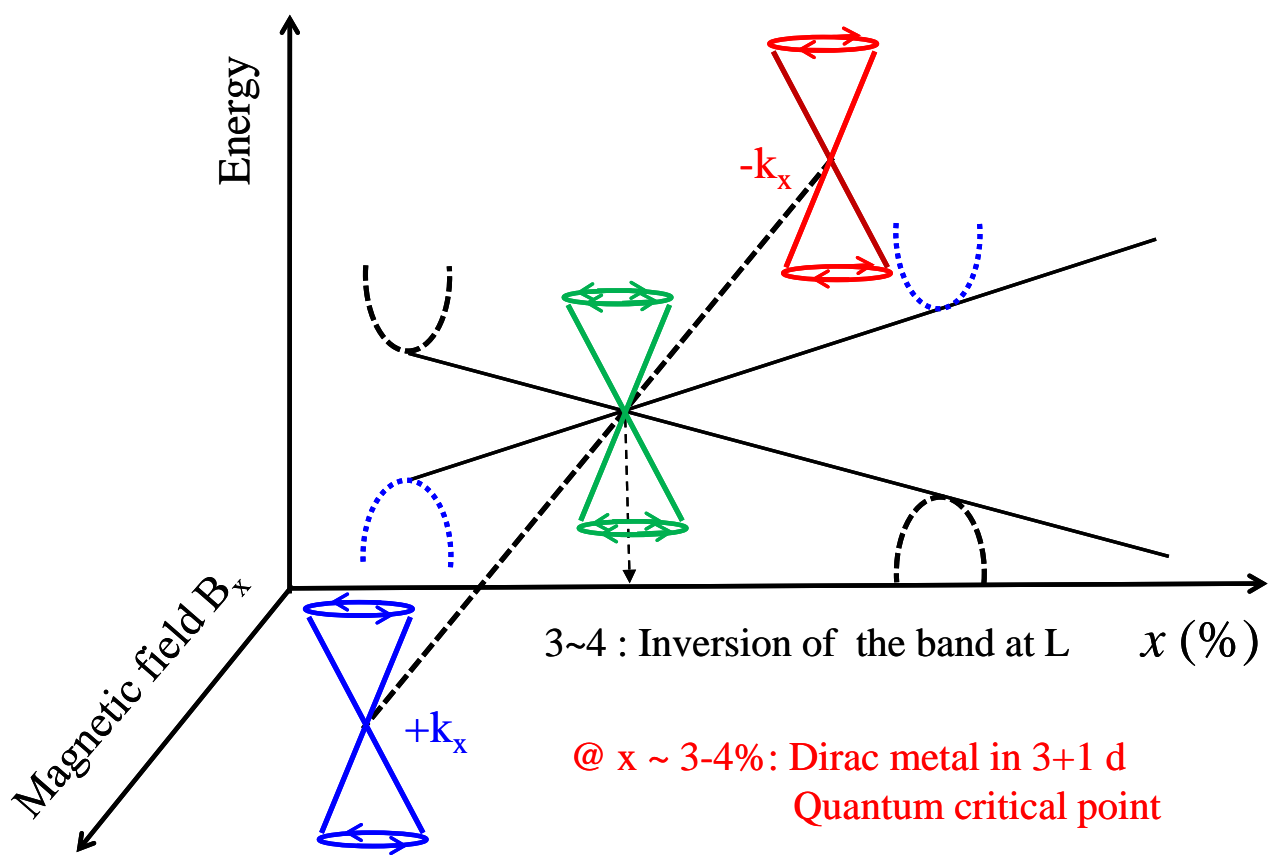

Figure 1(a) Schematic band structure of $Bi_{1-x}Sb_x$ near the L point

3~4 : Inversion of the band at L

@ x ~ 3-4%: Dirac metal in 3+1 d
Quantum critical point

Figure 1(b) **Measurements of anisotropic transport coefficients**

$$E_\alpha = \rho_{\alpha\beta}(B_\gamma) J_\beta$$

(1) $T$-MR, $\rho_{xx}(B_z)$      (2) $T$-Hall, $\rho_{yx}(B_z)$

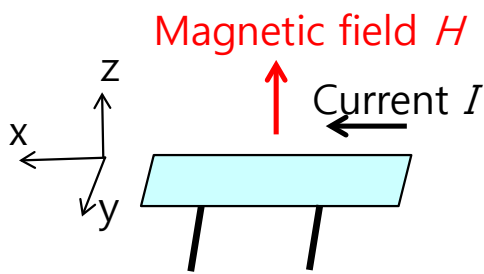
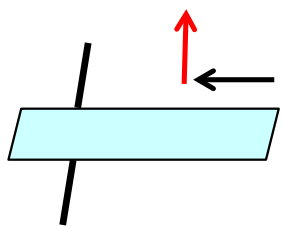

(3) $L_{//}$-MR, $\rho_{xx}(B_x)$      (4) $\rho_{yx}(B_x)$

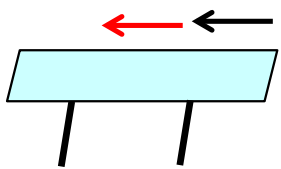
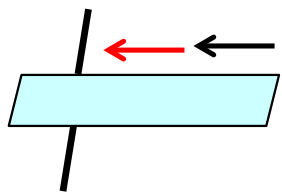

(5) $T'$-MR, $\rho_{xx}(B_y)$      (6) $\rho_{yx}(B_y)$

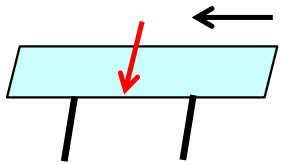
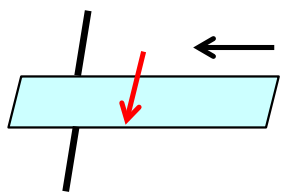

Figure 2

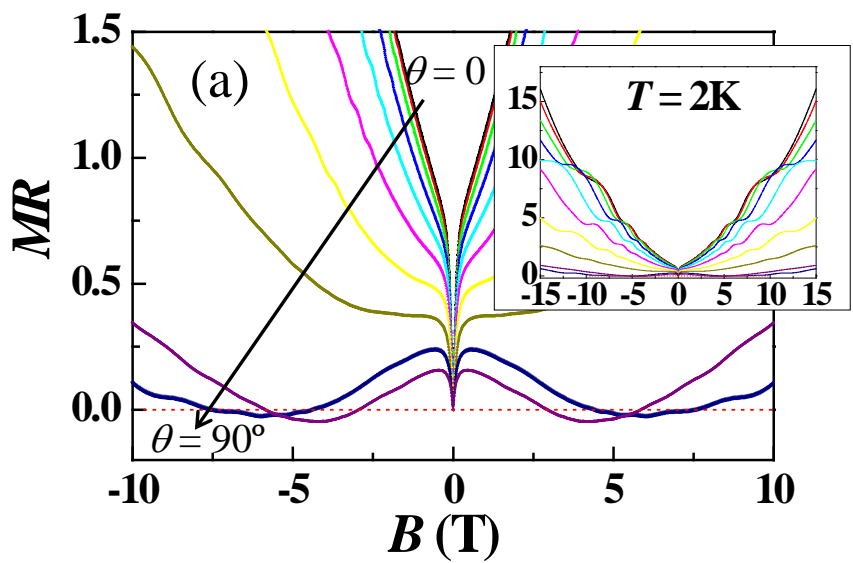

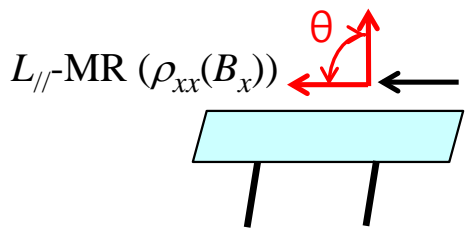

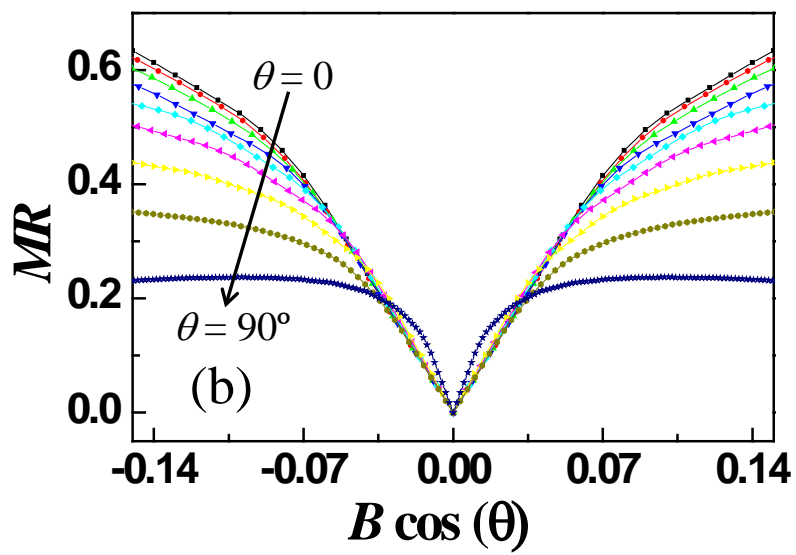

Figure 3

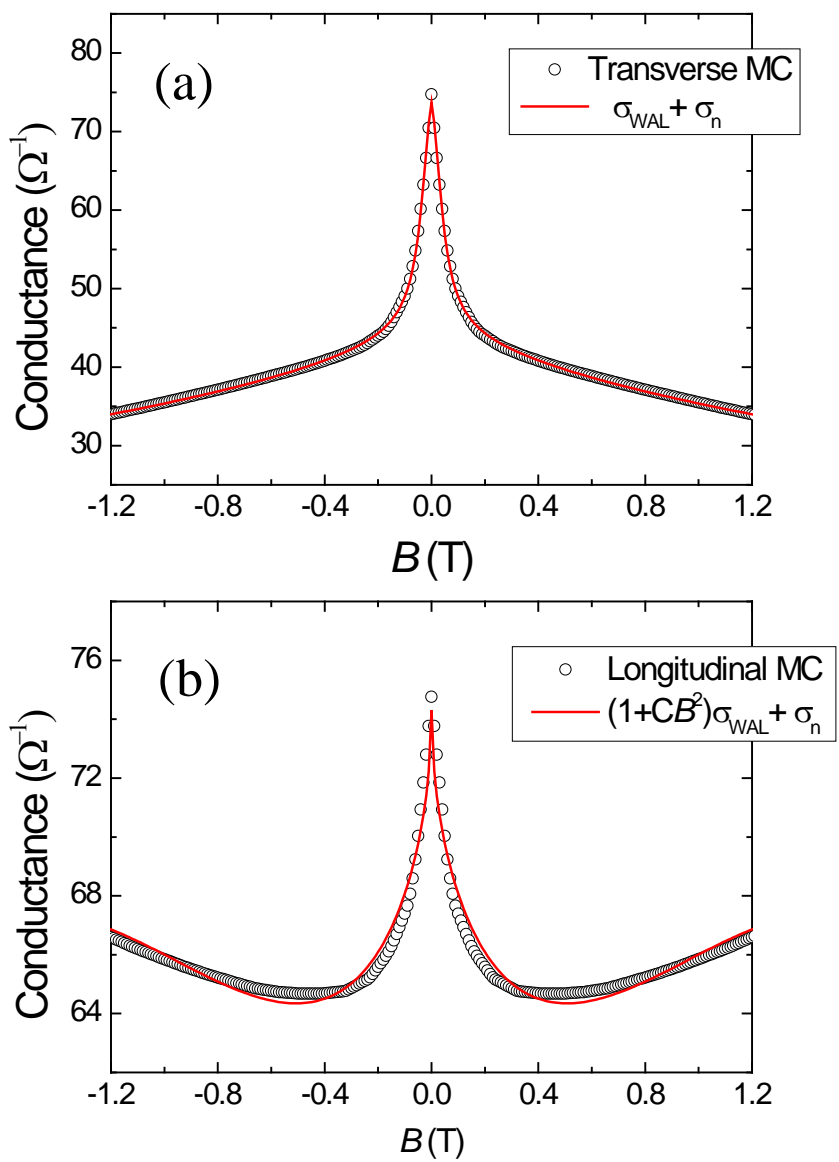



\title{A supplementary material for ``Dirac vs. Weyl in topological insulators: Adler-Bell-Jackiw anomaly in transport phenomena"}
\author{ Heon-Jung Kim$^{1}$, Ki-Seok Kim$^{2}$, J.-F. Wang$^{3}$,
M. Sasaki$^{4}$, N. Satoh$^{5}$, A. Ohnishi$^{4}$, M.
Kitaura$^{4}$, M. Yang$^{3}$, L. Li$^{3}$ } \affiliation{
$^{1}$Department of Physics, College of Natural
Science, Daegu University, Gyeongbuk 712-714, Republic of Korea \\
$^{2}$Department of Physics, Pohang University of Science and
Technology, Pohang, Gyeongbuk 790-784, Republic of Korea \\
$^{3}$Wuhan National High Magnetic Field Center, Huazhong
University of Science and Technology, Wuhan 430074, China \\
$^{4}$Department of Physics, Faculty of Science, Yamagata
University, Kojirakawa, Yamagata 990-8560 \\ $^{5}$Department of
Electronics and Computer Science, Iwaki Meisei University, Iwaki,
Fukushima 970-8551, Japan }
\date{\today}

\maketitle

\section{Adler-Bell-Jackiw anomaly and analysis of the experimental data
in the ultra-quantum limit}

\subsection{Adler-Bell-Jackiw anomaly}

The underlying mechanism for the upturn behavior of the longitudinal magnetoconductivity is the Adler-Bell-Jackiw anomaly, where the production of Weyl fermions with a given chirality, saying $+$ chirality, gives rise to the reduction of Weyl fermions with an opposite chirality, saying $-$ one. This can be translated into a transfer of electrons from one Weyl point with $-$ chirality to the other Weyl point with $+$ chirality. This dissipationless transport process should be balanced by either intra-node or inter-node (node = Weyl point) scattering due to impurities in order to reach a steady state. Associated with two types of scattering processes, one can introduce two time scales, $\tau_{intra}$ and $\tau_{inter}$, where $\tau_{intra}$ is the intra-scattering time due to impurities within the same Weyl cone while $\tau_{inter}$ is that between two paired Weyl points.

Although we focus on the semi-classical regime without Landau-level quantization, there are two regimes. The approach based on the semi-classical equations of motion is valid when $\mu \gg \hbar \omega_{c}$, where $\mu$ is the chemical potential of the system and $\omega_{c}$ is the cyclotron frequency. On the other hand, the quantum regime, which has been discussed intensively but rather intuitively in Refs. \cite{ABJ_Nielsen_Ninomiya,DTSon_Boltzmann,ABJ_Aji}, emerges in the limit of $\omega_{c} \gg 1/\tau_{tr}$. Here, $\tau_{tr} = \tau_{intra}$ is the transport time, where the contribution of back scattering is extracted from the relaxation time. In this quantum regime, one finds $\tau_{tr} = \tau_{intra} \rightarrow \infty \gg \tau_{inter}$ that results from effectively one-dimensional chiral dynamics of electrons due to Landau-level quantization. In the semi-classical regime, it is natural to assume $\tau_{intra} \ll \tau_{inter}$ because the inter-node scattering requires finite momentum-transfer \cite{DTSon_Boltzmann}. As a result, transport properties are basically determined by $\tau_{eff} = \frac{\tau_{intra} \tau_{inter}}{\tau_{intra} + \tau_{inter}} \approx \tau_{intra}$ in the semi-classical regime while it is $\tau_{inter}$, the short time-scale that governs electrical transport in the quantum regime.

In the semi-classical regime, the longitudinal magnetoconductivity is found to be $\sigma + \mbox{const. ~} B^{2} \sigma$, where $\sigma$ is the Drude conductivity proportional to the intra-node scattering time $\tau_{intra}$. The first term results from relaxation due to intra-node scattering processes and it exists even without the other Weyl point as expected. On the other hand, the second term originates from a transfer of electrons from one Weyl point ($-$ chirality) to the other Weyl point ($+$ chirality), where this dissipationless transfer is relaxed by the intra-node scattering, reaching a steady state. Since a finite momentum-transfer, proportional to the applied magnetic field, is involved in this transfer process, the $B^{2}$ contribution appears to enhance the longitudinal conductivity. In contrast, the condition of $\tau_{intra} \rightarrow \infty \gg \tau_{inter}$ in the quantum regime does not allow the intra-node scattering as the case of one-dimensional chiral fermions. Instead, the transport dynamics is governed by $\tau_{inter}$. As a result, the longitudinal magnetoconductivity is found to be $|B| \sigma'$, where $\sigma'$ is the conductivity associated with the inter-node scattering time, $\tau_{inter}$ \cite{ABJ_Nielsen_Ninomiya,DTSon_Boltzmann,ABJ_Aji}. There is no Drude-like term in this case and the linear $|B|$ dependence results from effectively one-dimensional chiral dynamics.

\subsection{Data analysis based on the formula of the ultra-quantum limit}

\subsubsection{Discussion on the ultra-quantum limit}

When strong magnetic fields are applied along the z-direction, the electron spectrum is given by $\epsilon_{n}(p_{z}) = \pm v \sqrt{2n (\hbar e/ c) B + p_{z}^{2}}$ with $n = 0, 1, 2, \ldots,$ where $v$ is the velocity of the Weyl fermion \cite{ABJ_Nielsen_Ninomiya,DTSon_Boltzmann}. In the lowest Landau level identified with $n = 0$, one obtains $\epsilon_{0} = \pm v p_{z}$, where $+$ and $-$ denote the right and left chirality of each Weyl point, respectively. The so called ultra-quantum limit means that both the chemical potential and temperature are less than the energy gap between the lowest and first Landau levels, i.e., $\mu, ~ T < \hbar v / L_{B}$, where $L_{B} = \sqrt{\hbar c / e B}$ is the magnetic length \cite{ABJ_Nielsen_Ninomiya,DTSon_Boltzmann}. Then, only chiral branches of the spectrum are occupied. This effectively one-dimensional dynamics (due to significant localization in the plane arising from strong magnetic fields) enhances the effect of the Adler-Bell-Jackiw anomaly, because the chirality of the lowest Landau level (ultra-quantum limit) causes the suppression of intra-Weyl-point scattering, originating from a lack of phase space. This is essentially the same as the case of one-dimensional chiral fermions. Contributions to the current from these one-dimensional chiral electrons can be relaxed only by inter-Weyl-points scattering processes, characterized by $\tau$.

Considering the corresponding semi-classical equations of motion $\dot{p}_{z} = e E_{z} - p_{z} / \tau$ and $v_{z} = \pm v$, Ref. \cite{DTSon_Boltzmann} finds \bqa && \sigma_{zz} = \frac{\tau e^{2} v}{4 \pi^{2} \hbar L_{B}^{2}} . \eqa An essential point of this expression is that the conductivity is linearly proportional to the applied magnetic field, distinguished from the $B^2$ dependence in the semi-classical regime without Landau-level quantization. In addition, intra-node (Weyl point) scattering is not allowed due to the one-dimensional chiral dynamics in the ultra-quantum limit.

\subsubsection{Data analysis in the ultra-quantum limit}

In order to check if our system is in the ultra-quantum limit,
we analyzed the experimental data based on the ultra-quantum-limit formula.
For fitting, we first replaced the one-dimensional inter-node scattering time, $\tau_{inter}$, with the scattering time associated with the one-dimensional weak anti-localization correction to reproduce the weak anti-localization part of the experimental data. We considered the formula of
$ \Bigl| B \Bigr| \sigma_{1D,WAL} + \sigma_n$
for the longitudinal magnetoconductivity, where $\sigma_{1D,WAL} = a/\sqrt{B}+\sigma_0$
and $\sigma_n = 1/(A B^2+\rho_0)$ are similarly defined as the case of the semi-classical regime. On the other hand, the transverse magnetoconductivity is quantitatively described by $ \sigma_{1D,WAL}^T + \sigma_n^T$, where $ \sigma_{1D,WAL}^T$ and $\sigma_n^T$ are corresponding quantities for the transverse configuration. We note that no contributions appear from the Adler-Bell-Jackiw anomaly in the transverse magnetoconductivity.
We fitted both the transverse and longitudinal magnetoconductivities based on the above formulae and carefully observed
the values of the fitting parameters for the consistency check.

Fig. S1 displays the experimental transverse magnetoconductivity up to $\pm 1.2$ T, fitted by two theoretical curves in the semi-classical regime [fig. S1(a)] and in the ultra-quantum limit [fig. S1(b)], respectively. Both theoretical curves simulated the experimental data quite reasonably, giving the fitting parameters of
$a = -14.3$ $\Omega^{-1}$T$^{-0.5}$, $\sigma_0 = 49.6$ $\Omega^{-1}$,
$\rho_0 = 4.15 \times 10^{-2}$ $\Omega$, and $A = 21.8$ $\Omega$ T$^{-2}$
for the semi-classical case and those of
$a = 4.15$ $\Omega^{-1}$T$^{0.5}$, $\sigma_0 = 22.6$ $\Omega^{-1}$,
$\rho_0 = 0.077$ $\Omega$, and $A = 0.037 $ $\Omega$ T$^{-2}$ for the ultra-quantum limit.
These values will be compared to those of the fitting to the longitudinal magnetoconductivity.

\begin{figure}[t]
\includegraphics[width=0.5\textwidth]{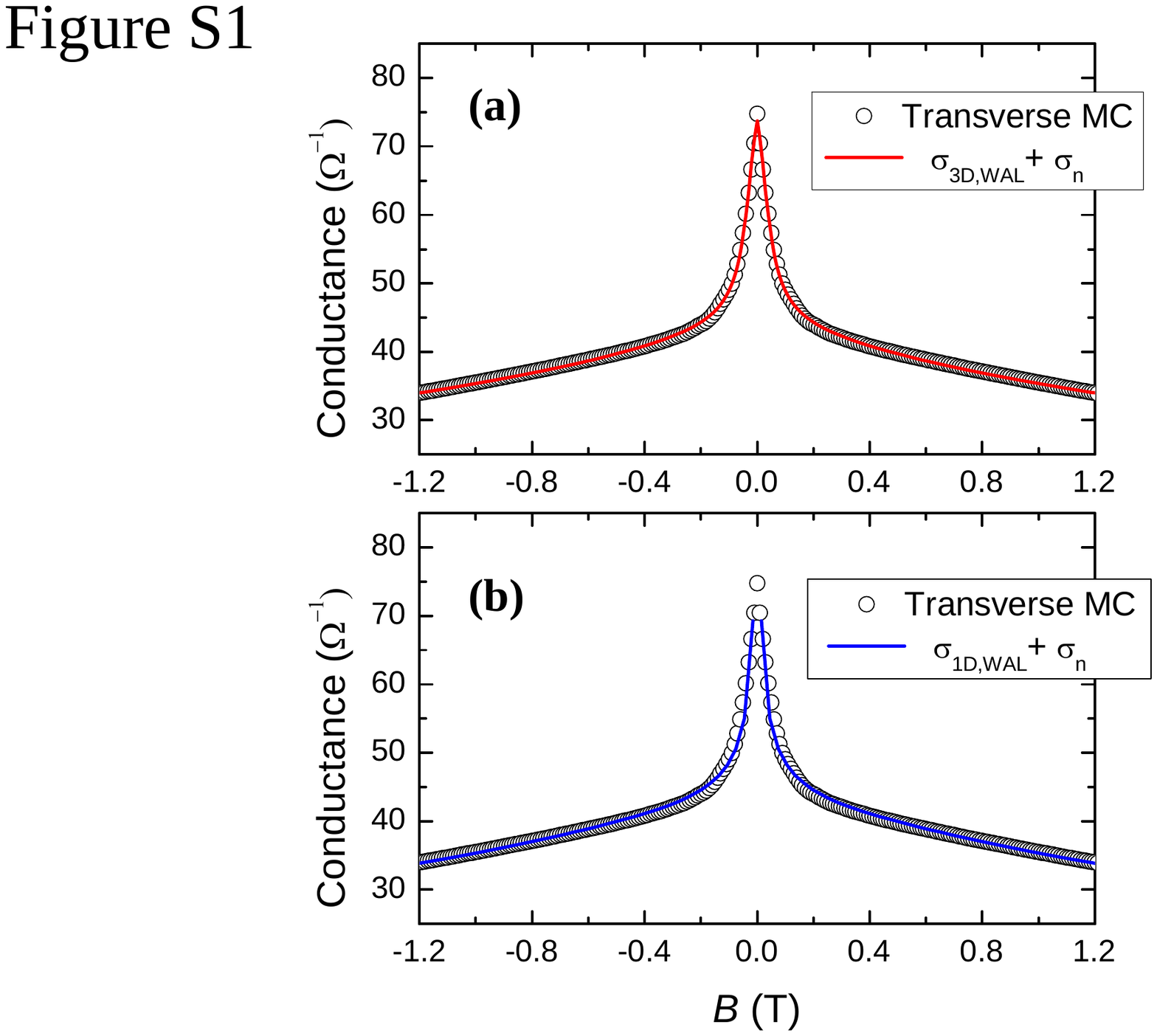}
\caption{Transverse magnetoconductivity with (a) theoretical fitting to the formula $\sigma_{3D,WAL}+\sigma_n$ in the semi-classical regime and (b) fitting to the formula $\sigma_{1D,WAL}+\sigma_n$ in the ultra-quantum limit.} \label{figS1}
\end{figure}

For the longitudinal configuration, the experimental data were fitted by the ultra-quantum-limit formula of
$ \Bigl| B \Bigr| \sigma_{1D,WAL} + \sigma_n$, modified from the correction of the Adler-Bell-Jackiw anomaly in the ultra-quantum limit. The fitting result to this formula, presented in Fig. S2(a), turns out to be quite poor. In this case, the fitting parameters are
$a = 4.15$ $\Omega^{-1}$T$^{0.5}$, $\sigma_0 = 3.67 \times 10^3$ $\Omega^{-1}$,
$\rho_0 = 0.0152$ $\Omega$, $C_W = 1.05 \times 10^{-4}$ T$^{-1}$,
 and $A \approx 0$ $\Omega$ T$^{-2}$.
In fact, we found that both the weak anti-localization correction and the upturn in the longitudinal magnetoconductivity cannot be fitted simultaneously as long as both $a$ and $A$ are positive as they should be.

We also investigated the possibility that the weak anti-localization correction and
the upturn behavior of the longitudinal magnetoconductivity arise in a separate way, where the former appears
from three-dimensional dynamics of electrons in the presence of the spin-orbit coupling and the latter comes from the effective one-dimensional chiral dynamics, respectively. In this case, the conductivity is expressed
by $\sigma_{3D}+\sigma_n+\Bigl| B \Bigr| C_W$, where $C_W$ is
associated with the inter-node scattering time. At this moment, we cannot clearly explain
the origin of the three-dimensional weak anti-localization correction. The fitting based on
this formula is presented in Fig. S2(b) with the fitting parameters of
$a = -13.4$ $\Omega^{-1}$T$^{-0.5}$, $\sigma_0 = 68.9$ $\Omega^{-1}$,
$\rho_0 = 0.165$ $\Omega$, $C_W=10.2$ $\Omega^{-1}$T$^{-1}$ and $A = 43.5$ $\Omega$ T$^{-2}$.
Compared to the transverse case, the value of $A$ increases significantly, which cannot be physical, because the orbital contribution from normal electrons must be absent in the longitudinal measurements.

All analysis support that our system is not near the ultra-quantum limit but in the semi-classical regime, which
is quite in agreement with the band structure of Bi$_{1-x}$Sb$_x$ at $x \sim 3$ $\%$, where the Fermi level is located far above the Weyl points.
Actually, this tendency associated with the chemical potential is quite consistent with anomalous transport physics of our previous studies \cite{TI_Ourexperiments}.

\begin{figure}[t]
\includegraphics[width=0.5\textwidth]{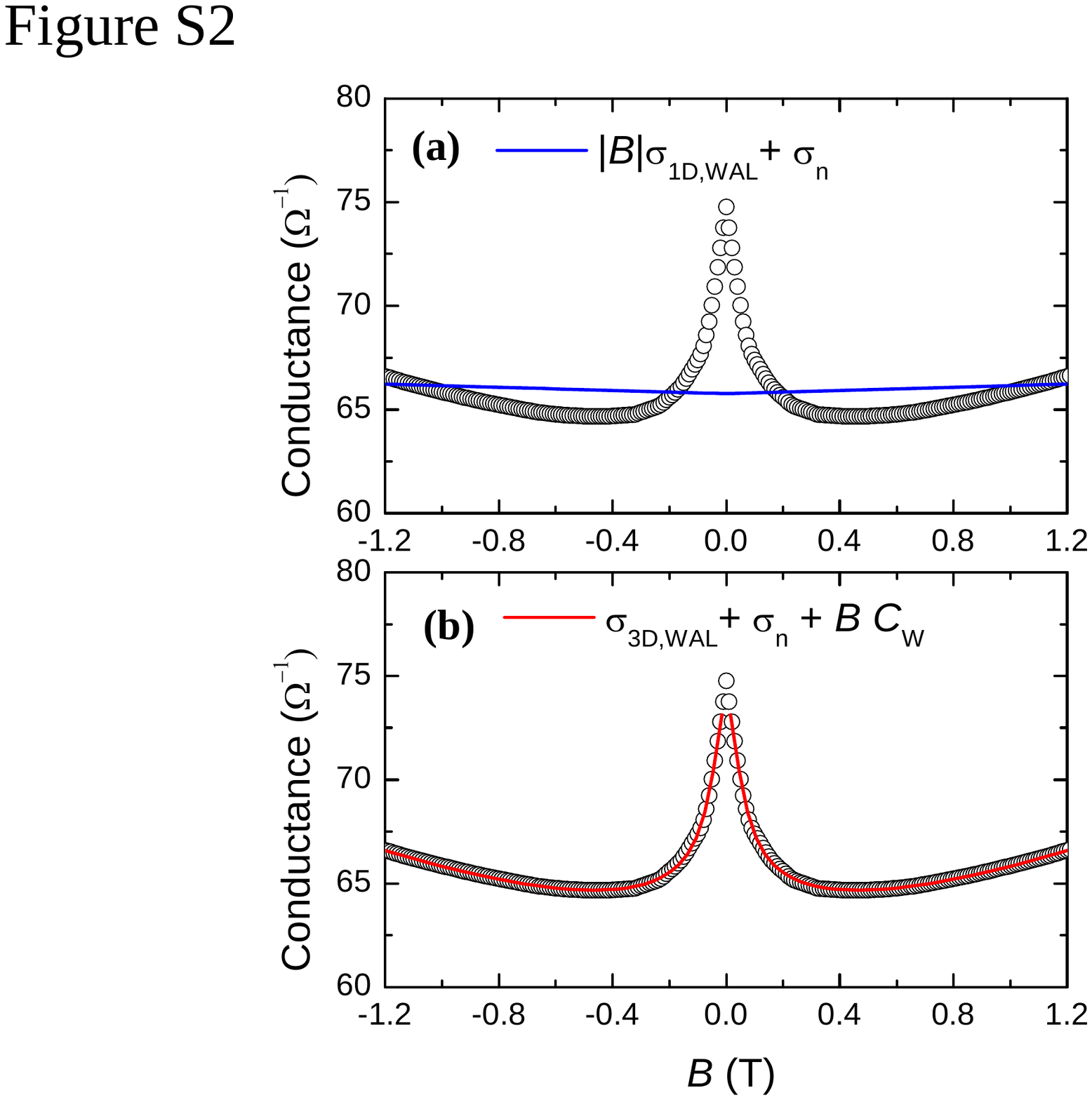}
\caption{Longitudinal magnetoconductivity with (a) theoretical fitting to the formula, $ \Bigl| B \Bigr| \sigma_{1D,WAL} + \sigma_n$ and (b) fitting to the formula,
$\sigma_{3D}+\sigma_n+\Bigl| B \Bigr| C_W$.} \label{figS2}
\end{figure}

\subsection{``Reentrant" downturn in the longitudinal magnetoconductivity above {\it B} $\sim$ 4 T}

The positive component of the longitudinal magnetoconductivity disappears eventually at the critical magnetic field of $B \sim 4$ T and the magnetoconductivity decreases with increasing $B$ above 4 T. One possible origin of this ``reentrant" downturn of magnetoconductivity can be found in the ``pair-annihilation" scenario, where a magnetic monopole (Weyl point with + chiral charge) recombines with an anti-monopole (Weyl point with $-$ chiral charge) to be annihilated around this particular magnetic field.
In fact, considering that the distance between the Weyl points is given by
$\sim 2\frac{gB}{v_D} \frac{e}{m^{\ast}}$, where  $g$ is the Lande  $g$-factor,
$v_D$  is the Dirac velocity at the critical point, $e$ is the charge of an electron,
and $m^{\ast}$  is the effective mass of the charge carrier, the critical magnetic field that corresponds to the largest distance in the Brillouin zone is estimated to be
$B_c \sim \frac{\pi}{a} \frac{v_D}{2g} \frac{m^{\ast}}{e}$.
Inserting the approximate values of $v_D  \sim 10^5$ m/s and  $g \sim$ 100
for topological insulators of BiSb alloys into this expression, we obtain $B_c \sim  1$ - 10 $T$, which resides within our experimental range.

However, a cautious person may criticize that it is unrealistic to consider such paired Weyl points to move all over the whole Brillouin zone away from the L point, because the band gap is rather large for the wave vectors away from the L point. Although we could exclude the possibility of Landau-level quantization in the region of weak magnetic fields associated with the Adler-Bell-Jackiw anomaly, it is difficult to exclude the possibility for the formation of Landau levels, increasing magnetic fields to reach the "intermediate" region for the reentrant downturn behavior. Indeed, Shubnikov-de Haas oscillations seem start right at the downturn field. We are suspecting that dominant contributions from higher Landau levels may cause this downturn behavior of magnetoconductivity in the intermediate region of magnetic fields. More concretely, it is natural to expect that the ultra-quantum limit is difficult to be achieved even in the region of such intermediate magnetic fields, because our samples are in the semi-classical regime originally, and thus higher Landau levels would be filled. Then, localization contributions may be dominant from higher Landau levels.

We believe that no theoretical studies have been performed for this crossover regime yet. This problem on the crossover regime is beyond the scope of the present investigation, which needs to be examined more sincerely near future both experimentally and theoretically.

\section{A formal development of the quantum Boltzmann equation in
the presence of the topological $\boldsymbol{E}\cdot\boldsymbol{B}$ term}

\subsection{Quantum Boltzmann equation}

We start from the quantum Boltzmann equation for a steady state
\cite{Mahan_Boltzmann} \bqa && \boldsymbol{\dot{p}} \cdot
\frac{\partial G^{<}(\boldsymbol{p},\omega)}{\partial
\boldsymbol{p}} + \boldsymbol{\dot{r}} \cdot \boldsymbol{\dot{p}}
\frac{\partial G^{<}(\boldsymbol{p},\omega)}{\partial \omega} -
\boldsymbol{\dot{p}} \cdot \Bigl\{ \frac{\partial
\Sigma^{<}(\boldsymbol{p},\omega)}{\partial \omega} \frac{\partial
\Re G_{ret}(\boldsymbol{p},\omega)}{\partial \boldsymbol{p}} -
\frac{\partial \Re G_{ret}(\boldsymbol{p},\omega)}{\partial
\omega} \frac{\partial \Sigma^{<}(\boldsymbol{p},\omega)}{\partial
\boldsymbol{p}} \Bigr\} \nn && = - 2 \Gamma(\boldsymbol{p},\omega)
G^{<}(\boldsymbol{p},\omega) + \Sigma^{<}(\boldsymbol{p},\omega)
A(\boldsymbol{p},\omega) . \eqa $G^{<}(\boldsymbol{p},\omega)$ is
the lesser Green's function, regarded as a quantum distribution
function, where $\boldsymbol{p}$ and $\omega$ represent momentum
and frequency for relative coordinates, respectively.
$\dot{\mathcal{O}}$ denotes the derivative with respect to time
$t$. $\Sigma^{<}(\boldsymbol{p},\omega)$ and
$G_{ret}(\boldsymbol{p},\omega)$ indicate the lesser self-energy
and the retarded Green's function, respectively, where $\Re$ is
their real part. The right hand side introduces collision terms,
where $\Gamma(\boldsymbol{p},\omega)$ and
$A(\boldsymbol{p},\omega)$ indicate the scattering rate and the
spectral function.

\subsection{Semi-classical equations of motion}

$\boldsymbol{r}$ and $\boldsymbol{p}$ are governed by
semi-classical equations of motion \cite{Berry_Phase}, given by
\bqa && \boldsymbol{\dot{r}} = \frac{\partial
\epsilon_{\boldsymbol{p}}}{\partial \boldsymbol{p}} +
\boldsymbol{\dot{p}} \times \boldsymbol{\Omega}_{\boldsymbol{p}} ,
\nn && \boldsymbol{\dot{p}} = e \boldsymbol{E} + \frac{e}{c}
\boldsymbol{\dot{r}} \times \boldsymbol{B} , \eqa where
$\boldsymbol{\Omega}_{\boldsymbol{p}}$ represents the Berry
curvature of the momentum space. Solving these equations, one
obtains \bqa && \boldsymbol{\dot{r}} = \Bigl( 1 + \frac{e}{c}
\boldsymbol{B} \cdot \boldsymbol{\Omega}_{\boldsymbol{p}}
\Bigr)^{-1} \Bigl\{ \boldsymbol{v}_{\boldsymbol{p}} + e
\boldsymbol{E} \times \boldsymbol{\Omega}_{\boldsymbol{p}} +
\frac{e}{c} \boldsymbol{\Omega}_{\boldsymbol{p}} \cdot
\boldsymbol{v}_{\boldsymbol{p}} \boldsymbol{B} \Bigr\} , \nn &&
\boldsymbol{\dot{p}} = \Bigl( 1 + \frac{e}{c} \boldsymbol{B} \cdot
\boldsymbol{\Omega}_{\boldsymbol{p}} \Bigr)^{-1} \Bigl\{ e
\boldsymbol{E} + \frac{e}{c} \boldsymbol{v}_{\boldsymbol{p}}
\times \boldsymbol{B} + \frac{e^{2}}{c} (\boldsymbol{E} \cdot
\boldsymbol{B}) \boldsymbol{\Omega}_{\boldsymbol{p}} \Bigr\} .
\eqa An essential point is the presence of the $\boldsymbol{E}
\cdot \boldsymbol{B}$ term in the second equation, imposing the
Adler-Bell-Jackiw anomaly \cite{DTSon_Boltzmann}.

A cautious person may ask how these semi-classical equations reflect dynamics of Weyl electrons in three dimensions because the equation of $\boldsymbol{\dot{p}}$ seems to apply to generic systems with non-zero Berry curvature, and not limited to Weyl metallic phases. Generally speaking, we have $\boldsymbol{\Omega}_{\boldsymbol{p}} = \boldsymbol{\Omega}_{-\boldsymbol{p}}$ in systems of time reversal symmetry while we get $\boldsymbol{\Omega}_{\boldsymbol{p}} = - \boldsymbol{\Omega}_{-\boldsymbol{p}}$ in centro-symmetric systems \cite{DTSon_Boltzmann}. As a result, we cannot avoid to reach the conclusion $\boldsymbol{\Omega}_{\boldsymbol{p}} = 0$ when systems contain both time reversal and inversion symmetries. Since the Dirac system contains both time reversal and inversion symmetries, we reach the conclusion that the Berry curvature vanishes, inevitably.

More mathematically speaking, the Berry curvature is given by the so called chiral charge, regarded as a topological charge (invariant). The Dirac band has both $+$ (right) and $-$ (left) chiral charges, and their sum give zero, corresponding to the vanishing Berry curvature. In other words, a magnetic monopole with $+$ charge exists together with that with $-$ charge at the same momentum point, which does not allow a non-zero Berry curvature. On the other hand, each Weyl point is characterized by each chiral charge, either $+$ or $-$, which should appear in a pair. As a result, each Weyl point, identified with a magnetic monopole of its corresponding charge, gives rise to a non-zero Berry curvature.

\subsection{Self-energy correction from intra-node scattering}

Inserting these equations into the quantum Boltzmann equation and
performing some algebra, we obtain the following expression \bqa
&& \Bigl( 1 + \frac{e}{c} \boldsymbol{B} \cdot
\boldsymbol{\Omega}_{\boldsymbol{p}} \Bigr)^{-1} \frac{e}{c}
\boldsymbol{v}_{\boldsymbol{p}} \cdot \Bigl( \boldsymbol{B} \times
\frac{\partial G^{<}}{\partial \boldsymbol{p}} \Bigr) + \Bigl( 1 +
\frac{e}{c} \boldsymbol{B} \cdot
\boldsymbol{\Omega}_{\boldsymbol{p}} \Bigr)^{-2} \Bigl(
\frac{e}{c} \boldsymbol{v}_{\boldsymbol{p}} \times \boldsymbol{B}
\Bigr) \cdot ( e \boldsymbol{E} \times
\boldsymbol{\Omega}_{\boldsymbol{p}} ) \frac{\partial
G^{<}}{\partial \omega} \nn && - [A(\boldsymbol{p},\omega)]^{2}
\Bigl( - \frac{\partial f(\omega)}{\partial \omega} \Bigr) \Bigl(
1 + \frac{e}{c} \boldsymbol{B} \cdot
\boldsymbol{\Omega}_{\boldsymbol{p}} \Bigr)^{-2} \Bigl\{ e
\boldsymbol{E} + \frac{e^{2}}{c} (\boldsymbol{E} \cdot
\boldsymbol{B}) \boldsymbol{\Omega}_{\boldsymbol{p}} \Bigr\} \cdot
\Bigl\{ \boldsymbol{v}_{\boldsymbol{p}} + \frac{e}{c}
(\boldsymbol{\Omega}_{\boldsymbol{p}} \cdot
\boldsymbol{v}_{\boldsymbol{p}}) \boldsymbol{B} \Bigr\} \Gamma \nn
&& = - i [ 2 \Gamma G^{<} - \Sigma^{<} A ] , \eqa where the
argument of $(\boldsymbol{p},\omega)$ is omitted for simplicity.

The lesser self-energy is given by \bqa &&
\Sigma^{<}(\boldsymbol{p},\omega) = \sum_{\boldsymbol{q}}
\int_{0}^{\infty} d \nu \Bigl| \frac{\boldsymbol{p} \times
\boldsymbol{\hat{q}}}{m} \Bigr|^{2} \Im D_{a}(\boldsymbol{q},\nu)
\Bigl\{ [n(\nu) + 1]
G^{<}(\boldsymbol{p}+\boldsymbol{q},\omega+\nu) + n(\nu)
G^{<}(\boldsymbol{p}+\boldsymbol{q},\omega-\nu) \Bigr\} . \nn \eqa
Here, we consider gauge interactions \cite{KiSeok_Boltzmann} for example. Thus,
$D_{a}(\boldsymbol{q},\nu)$ represents the Green function of gauge
fluctuations. $n(\nu)$ is the Bose-Einstein distribution function.
One can replace the gauge-boson propagator with some other types
of fluctuations such as phonons, spin fluctuations, and etc. One
may consider the diffusion-mode propagator for weak
anti-localization in the presence of spin-orbit coupling, where
the form of its vertex should be changed, of course. Although this disorder scattering will be
most relevant in the present problem, the precise form of the lesser self-energy is not important for our derivation.

\subsection{Ansatz for the lesser Green's function}

We write down the lesser Green's function in the following way
\cite{KiSeok_Boltzmann} \bqa && G^{<}(\boldsymbol{p},\omega) = i
f(\omega) A(\boldsymbol{p},\omega) + i \Bigl( - \frac{\partial
f(\omega)}{\partial \omega} \Bigr) A(\boldsymbol{p},\omega)
\boldsymbol{v}_{\boldsymbol{p}} \cdot
\Lambda(\boldsymbol{p},\omega) , \eqa which consists of the
equilibrium part (the first term) and its correction term (the
second term). We call $\Lambda(\boldsymbol{p},\omega)$ ``vertex
distribution-function" although it sounds somewhat confusing.
$f(\omega)$ is the Fermi-Dirac distribution function.

\subsection{Scattering rate}

Inserting this ansatz into the quantum Boltzmann equation with the
expression of the lesser self-energy and performing some
straightforward algebra, we obtain \bqa && i \Bigl( 1 +
\frac{e}{c} \boldsymbol{B} \cdot
\boldsymbol{\Omega}_{\boldsymbol{p}} \Bigr)^{-1} \frac{e}{m c}
\boldsymbol{v}_{\boldsymbol{p}} \cdot \Bigl( \boldsymbol{B} \times
\frac{\partial \boldsymbol{p}_{\alpha}}{\partial \boldsymbol{p}}
\Bigr) \Lambda_{\alpha}(\boldsymbol{p},\omega) - i \Bigl( 1 +
\frac{e}{c} \boldsymbol{B} \cdot
\boldsymbol{\Omega}_{\boldsymbol{p}} \Bigr)^{-2} \Bigl(
\frac{e}{c} \boldsymbol{v}_{\boldsymbol{p}} \times \boldsymbol{B}
\Bigr) \cdot ( e \boldsymbol{E} \times
\boldsymbol{\Omega}_{\boldsymbol{p}} ) \nn && -
A(\boldsymbol{p},\omega) \Bigl( 1 + \frac{e}{c} \boldsymbol{B}
\cdot \boldsymbol{\Omega}_{\boldsymbol{p}} \Bigr)^{-2} \Bigl\{ e
\boldsymbol{E} + \frac{e^{2}}{c} (\boldsymbol{E} \cdot
\boldsymbol{B}) \boldsymbol{\Omega}_{\boldsymbol{p}} \Bigr\} \cdot
\Bigl\{ \boldsymbol{v}_{\boldsymbol{p}} + \frac{e}{c}
(\boldsymbol{\Omega}_{\boldsymbol{p}} \cdot
\boldsymbol{v}_{\boldsymbol{p}}) \boldsymbol{B} \Bigr\}
\Gamma(\boldsymbol{p},\omega) \nn && = 2
\Gamma(\boldsymbol{p},\omega) \boldsymbol{v}_{\boldsymbol{p}}
\cdot \Lambda(\boldsymbol{p},\omega) - \sum_{\boldsymbol{q}}
\int_{0}^{\infty} d \nu \Bigl| \frac{\boldsymbol{p} \times
\boldsymbol{\hat{q}}}{m} \Bigr|^{2} \Im D_{a}(\boldsymbol{q},\nu) \nn && \times
\Bigl\{ [n(\nu) + f(\omega+\nu)]
A(\boldsymbol{p}+\boldsymbol{q},\omega+\nu)
\boldsymbol{v}_{\boldsymbol{p}+\boldsymbol{q}} \cdot
\Lambda(\boldsymbol{p}+\boldsymbol{q},\omega+\nu) \nn && -
[n(-\nu) + f(\omega-\nu)]
A(\boldsymbol{p}+\boldsymbol{q},\omega-\nu)
\boldsymbol{v}_{\boldsymbol{p}+\boldsymbol{q}} \cdot
\Lambda(\boldsymbol{p}+\boldsymbol{q},\omega-\nu) \Bigr\} , \eqa
where we have used the following relation \bqa && 2
\Gamma(\boldsymbol{p},\omega) = \sum_{\boldsymbol{q}}
\int_{0}^{\infty} d \nu \Bigl| \frac{\boldsymbol{p} \times
\boldsymbol{\hat{q}}}{m} \Bigr|^{2} \Im D_{a}(\boldsymbol{q},\nu)
\Bigl\{ [ n(\nu) + f(\omega+\nu) ]
A(\boldsymbol{p}+\boldsymbol{q},\omega+\nu) \nn && - [ n(-\nu) + f(\omega
- \nu) ] A(\boldsymbol{p}+\boldsymbol{q},\omega-\nu) \Bigr\} .
\eqa

\subsection{Vertex distribution-function}

Writing down the quantum Boltzmann equation in terms of components
and focusing on dynamics near the Fermi surface, we reach the
following expression for each component, \bqa &&
\frac{\Lambda_{F}^{x}(\omega)}{\tau_{tr}(\omega)} + i \Bigl( 1 +
\frac{e}{c} \boldsymbol{B} \cdot \boldsymbol{\Omega}_{F}
\Bigr)^{-1} \frac{e B_{z}}{m c} \Lambda_{F}^{y}(\omega) - i \Bigl(
1 + \frac{e}{c} \boldsymbol{B} \cdot \boldsymbol{\Omega}_{F}
\Bigr)^{-1} \frac{e B_{y}}{m c} \Lambda_{F}^{z}(\omega) \nn && = -
i \Bigl( 1 + \frac{e}{c} \boldsymbol{B} \cdot
\boldsymbol{\Omega}_{F} \Bigr)^{-2} \Bigl\{ \frac{e}{c} B_{y} ( e
\boldsymbol{E} \times \boldsymbol{\Omega}_{F} )_{z} - \frac{e}{c}
B_{z} ( e \boldsymbol{E} \times \boldsymbol{\Omega}_{F} )_{y}
\Bigr\} \nn && - A(\boldsymbol{p}_{F},\omega)
\Gamma(\boldsymbol{p}_{F},\omega) \Bigl( 1 + \frac{e}{c}
\boldsymbol{B} \cdot \boldsymbol{\Omega}_{F} \Bigr)^{-2} \Bigl\{ e
E_{x} + \frac{e^{2}}{c} (\boldsymbol{E} \cdot \boldsymbol{B})
\boldsymbol{\Omega}_{F}^{x} + \frac{e^{2}}{c} \Bigl( 1 +
\frac{e}{c} \boldsymbol{B} \cdot \boldsymbol{\Omega}_{F} \Bigr)
(\boldsymbol{E} \cdot \boldsymbol{B}) \boldsymbol{\Omega}_{F}^{x}
\Bigr\} , \nn \eqa \bqa &&
\frac{\Lambda_{F}^{y}(\omega)}{\tau_{tr}(\omega)} - i \Bigl( 1 +
\frac{e}{c} \boldsymbol{B} \cdot \boldsymbol{\Omega}_{F}
\Bigr)^{-1} \frac{e B_{z} }{m c} \Lambda_{F}^{x}(\omega) + i
\Bigl( 1 + \frac{e}{c} \boldsymbol{B} \cdot
\boldsymbol{\Omega}_{F} \Bigr)^{-1} \frac{e B_{x}}{m c}
\Lambda_{F}^{z}(\omega) \nn && = - i \Bigl( 1 + \frac{e}{c}
\boldsymbol{B} \cdot \boldsymbol{\Omega}_{F} \Bigr)^{-2} \Bigl\{ -
\frac{e}{c} B_{x} ( e \boldsymbol{E} \times
\boldsymbol{\Omega}_{F} )_{z} + \frac{e}{c} B_{z} ( e
\boldsymbol{E} \times \boldsymbol{\Omega}_{F} )_{x} \Bigr\} \nn &&
- A(\boldsymbol{p}_{F},\omega) \Gamma(\boldsymbol{p}_{F},\omega)
\Bigl( 1 + \frac{e}{c} \boldsymbol{B} \cdot
\boldsymbol{\Omega}_{F} \Bigr)^{-2} \Bigl\{ e E_{y} +
\frac{e^{2}}{c} (\boldsymbol{E} \cdot \boldsymbol{B})
\boldsymbol{\Omega}_{F}^{y} + \frac{e^{2}}{c} \Bigl( 1 +
\frac{e}{c} \boldsymbol{B} \cdot \boldsymbol{\Omega}_{F} \Bigr)
(\boldsymbol{E} \cdot \boldsymbol{B}) \boldsymbol{\Omega}_{F}^{y}
\Bigr\} , \nn \eqa and \bqa &&
\frac{\Lambda_{F}^{z}(\omega)}{\tau_{tr}(\omega)} + i \Bigl( 1 +
\frac{e}{c} \boldsymbol{B} \cdot \boldsymbol{\Omega}_{F}
\Bigr)^{-1} \frac{e B_{y} }{m c} \Lambda_{F}^{x}(\omega) - i
\Bigl( 1 + \frac{e}{c} \boldsymbol{B} \cdot
\boldsymbol{\Omega}_{F} \Bigr)^{-1} \frac{e B_{x}}{m c}
\Lambda_{F}^{y}(\omega) \nn && = - i \Bigl( 1 + \frac{e}{c}
\boldsymbol{B} \cdot \boldsymbol{\Omega}_{F} \Bigr)^{-2} \Bigl\{ -
\frac{e}{c} B_{y} ( e \boldsymbol{E} \times
\boldsymbol{\Omega}_{F} )_{x} + \frac{e}{c} B_{x} ( e
\boldsymbol{E} \times \boldsymbol{\Omega}_{F} )_{y} \Bigr\} \nn &&
- A(\boldsymbol{p}_{F},\omega) \Gamma(\boldsymbol{p}_{F},\omega)
\Bigl( 1 + \frac{e}{c} \boldsymbol{B} \cdot
\boldsymbol{\Omega}_{F} \Bigr)^{-2} \Bigl\{ e E_{z} +
\frac{e^{2}}{c} (\boldsymbol{E} \cdot \boldsymbol{B})
\boldsymbol{\Omega}_{F}^{z} + \frac{e^{2}}{c} \Bigl( 1 +
\frac{e}{c} \boldsymbol{B} \cdot \boldsymbol{\Omega}_{F} \Bigr)
(\boldsymbol{E} \cdot \boldsymbol{B}) \boldsymbol{\Omega}_{F}^{z}
\Bigr\} , \nn \eqa where the transport time is given by \bqa &&
\frac{1}{\tau_{tr}(\omega)} = \sum_{\boldsymbol{q}}
\int_{0}^{\infty} d \nu \Bigl| \frac{\boldsymbol{p}_{F} \times
\boldsymbol{\hat{q}}}{m} \Bigr|^{2} \Im D_{a}(\boldsymbol{q},\nu)
(1 -  \cos \theta) \Bigl\{ [n(\nu) + f(\omega+\nu)]
A(\boldsymbol{p}_{F}+\boldsymbol{q},\omega+\nu) \nn && - [n(-\nu)
+ f(\omega-\nu)] A(\boldsymbol{p}_{F}+\boldsymbol{q},\omega-\nu)
\Bigr\} . \eqa We note the $1 - \cos \theta$ factor in this
expression, which extracts out back scattering contributions.

\section{Current formulation}

It is natural to define a current in the following way
\cite{DTSon_Boltzmann} \bqa && \boldsymbol{J} = - e
\frac{1}{\beta} \sum_{i\omega} \int \frac{d^{3}
\boldsymbol{p}}{(2\pi \hbar)^{3}} \Bigl( 1 + \frac{e}{c}
\boldsymbol{B} \cdot \boldsymbol{\Omega}_{\boldsymbol{p}}
\Bigr)^{-1} \Bigl\{ \boldsymbol{v}_{\boldsymbol{p}} + e
\boldsymbol{E} \times \boldsymbol{\Omega}_{\boldsymbol{p}} +
\frac{e}{c} (\boldsymbol{\Omega}_{\boldsymbol{p}} \cdot
\boldsymbol{v}_{\boldsymbol{p}}) \boldsymbol{B} \Bigr\}[ - i
G^{<}(\boldsymbol{p},i\omega)] . \nn \eqa We note the
$\dot{\boldsymbol{r}}$ term in the integral expression.

Inserting the ansatz for the lesser Green's function into the
above expression, we obtain \bqa && \boldsymbol{J} = - e^{2}
\frac{1}{\beta} \sum_{i\omega} \int \frac{d^{3}
\boldsymbol{p}}{(2\pi \hbar)^{3}} \Bigl( 1 + \frac{e}{c}
\boldsymbol{B} \cdot \boldsymbol{\Omega}_{\boldsymbol{p}}
\Bigr)^{-1} (\boldsymbol{E} \times
\boldsymbol{\Omega}_{\boldsymbol{p}}) f(\omega)
A(\boldsymbol{p},\omega) \nn && - e \frac{1}{\beta} \sum_{i\omega}
\int \frac{d^{3} \boldsymbol{p}}{(2\pi \hbar)^{3}} \Bigl( 1 +
\frac{e}{c} \boldsymbol{B} \cdot
\boldsymbol{\Omega}_{\boldsymbol{p}} \Bigr)^{-1} \Bigl\{
\boldsymbol{v}_{\boldsymbol{p}} + \frac{e}{c}
(\boldsymbol{\Omega}_{\boldsymbol{p}} \cdot
\boldsymbol{v}_{\boldsymbol{p}}) \boldsymbol{B} \Bigr\} \Bigl( -
\frac{\partial f(\omega)}{\partial \omega} \Bigr)
A(\boldsymbol{p},\omega) \boldsymbol{v}_{\boldsymbol{p}} \cdot
\Lambda(\boldsymbol{p},\omega) . \nn \eqa Then, the $x-$component is
given by \bqa && J_{x} = - e^{2} \frac{1}{\beta} \sum_{i\omega}
\int \frac{d^{3} \boldsymbol{p}}{(2\pi \hbar)^{3}} \Bigl( 1 +
\frac{e}{c} \boldsymbol{B} \cdot
\boldsymbol{\Omega}_{\boldsymbol{p}} \Bigr)^{-1} (E_{y}
\boldsymbol{\Omega}_{\boldsymbol{p}}^{z} - E_{z}
\boldsymbol{\Omega}_{\boldsymbol{p}}^{y}) f(\omega)
A(\boldsymbol{p},\omega) \nn && - e \frac{1}{\beta} \sum_{i\omega}
\int \frac{d^{3} \boldsymbol{p}}{(2\pi \hbar)^{3}} \Bigl( 1 +
\frac{e}{c} \boldsymbol{B} \cdot
\boldsymbol{\Omega}_{\boldsymbol{p}} \Bigr)^{-1} (v_{F}^{x})^{2}
\Bigl( - \frac{\partial f(\omega)}{\partial \omega} \Bigr)
A(\boldsymbol{p},\omega) \Lambda_{x}(\boldsymbol{p},\omega) \nn &&
- e \frac{1}{\beta} \sum_{i\omega} \int \frac{d^{3}
\boldsymbol{p}}{(2\pi \hbar)^{3}} \Bigl( 1 + \frac{e}{c}
\boldsymbol{B} \cdot \boldsymbol{\Omega}_{\boldsymbol{p}}
\Bigr)^{-1} \frac{e}{c} B_{x} \Bigl\{ (v_{\boldsymbol{p}}^{x})^{2}
\boldsymbol{\Omega}_{\boldsymbol{p}}^{x}
\Lambda_{x}(\boldsymbol{p},\omega) + (v_{\boldsymbol{p}}^{y})^{2}
\boldsymbol{\Omega}_{\boldsymbol{p}}^{y}
\Lambda_{y}(\boldsymbol{p},\omega) \nn && + (v_{\boldsymbol{p}}^{z})^{2}
\boldsymbol{\Omega}_{\boldsymbol{p}}^{z}
\Lambda_{z}(\boldsymbol{p},\omega) \Bigr\} \Bigl( - \frac{\partial
f(\omega)}{\partial \omega} \Bigr) A(\boldsymbol{p},\omega) .
\eqa

\section{Anomalous transport phenomena}

\subsection{$\boldsymbol{B} = B_{z} \boldsymbol{\hat{z}}$ and $\boldsymbol{E} = E_{x} \boldsymbol{\hat{x}}$}

Solving the quantum Boltzmann equation in the conventional setup
of $\boldsymbol{B} = B_{z} \boldsymbol{\hat{z}}$ and
$\boldsymbol{E} = E_{x} \boldsymbol{\hat{x}}$, we find \bqa &&
\Lambda_{F}^{x}(\omega) \approx - e
\frac{A(\boldsymbol{p}_{F},\omega)
\frac{\tau_{tr}(\omega)}{\tau_{sc}(\omega)} E_{x} }{\Bigl( 1 +
\frac{e}{c} B_{z} \boldsymbol{\Omega}_{F}^{z} \Bigr)^{2} +
[\omega_{c}^{z} \tau_{tr}(\omega)]^{2} } - \frac{m e}{1 +
\frac{e}{c} B_{z} \boldsymbol{\Omega}_{F}^{z}}
\frac{\boldsymbol{\Omega}_{F}^{z} [\omega_{c}^{z}
\tau_{tr}(\omega)]^{2} E_{x} }{\Bigl( 1 + \frac{e}{c} B_{z}
\boldsymbol{\Omega}_{F}^{z} \Bigr)^{2} + [\omega_{c}^{z}
\tau_{tr}(\omega)]^{2} } \eqa and \bqa && \Lambda_{F}^{y}(\omega)
= \frac{e}{1 + \frac{e}{c} B_{z} \boldsymbol{\Omega}_{F}^{z}}
\frac{A(\boldsymbol{p}_{F},\omega)
\frac{\tau_{tr}(\omega)}{\tau_{sc}(\omega)} [\omega_{c}^{z}
\tau_{tr}(\omega)] E_{x} }{\Bigl( 1 + \frac{e}{c} B_{z}
\boldsymbol{\Omega}_{F}^{z} \Bigr)^{2} + [\omega_{c}^{z}
\tau_{tr}(\omega)]^{2} } - m e \frac{\boldsymbol{\Omega}_{F}^{z}
[\omega_{c}^{z} \tau_{tr}(\omega)] E_{x}}{\Bigl( 1 + \frac{e}{c}
B_{z} \boldsymbol{\Omega}_{F}^{z} \Bigr)^{2} + [\omega_{c}^{z}
\tau_{tr}(\omega)]^{2} } \eqa with \bqa &&
\frac{\Lambda_{F}^{z}(\omega)}{\tau_{tr}(\omega)} = 0 . \eqa Here,
we replaced $\Gamma(\boldsymbol{p}_{F},\omega)$ with
$\frac{1}{\tau_{sc}(\omega)}$. $\omega_{c}^{z} = \frac{e B_{z}}{m
c}$ is the cyclotron frequency.

Inserting the vertex distribution-functions into the current
formula, we obtain \bqa && J_{x} \approx e^{2} N_{F}
\frac{1}{\beta} \sum_{i\omega} \int_{-\infty}^{\infty} d \epsilon
\int_{-1}^{1} d \cos \theta \int_{0}^{2\pi} d \phi (v_{F}^{x})^{2}
\Bigl( - \frac{\partial f(\omega)}{\partial \omega} \Bigr)
[A_{F}(\epsilon,\omega)]^{2}
\frac{\frac{\tau_{tr}(\omega)}{\tau_{sc}(\omega)} }{1 +
[\omega_{c}^{z} \tau_{tr}(\omega)]^{2} } E_{x} \nn && +
\frac{e^{4}}{c^{2}} N_{F} \frac{1}{\beta} \sum_{i\omega}
\int_{-\infty}^{\infty} d \epsilon \int_{-1}^{1} d \cos \theta
\int_{0}^{2\pi} d \phi (v_{F}^{x})^{2}
(\boldsymbol{\Omega}_{F}^{z})^{2} \Bigl( - \frac{\partial
f(\omega)}{\partial \omega} \Bigr) [A_{F}(\epsilon,\omega)]^{2}
\frac{\frac{\tau_{tr}(\omega)}{\tau_{sc}(\omega)} }{1 +
[\omega_{c}^{z} \tau_{tr}(\omega)]^{2} } B_{z}^{2} E_{x} \nn && -
2 m \frac{e^{3}}{c} N_{F} \frac{1}{\beta} \sum_{i\omega}
\int_{-\infty}^{\infty} d \epsilon \int_{-1}^{1} d \cos \theta
\int_{0}^{2\pi} d \phi (v_{F}^{x})^{2}
(\boldsymbol{\Omega}_{F}^{z})^{2} \Bigl( - \frac{\partial
f(\omega)}{\partial \omega} \Bigr) A_{F}(\epsilon,\omega)
\frac{[\omega_{c}^{z} \tau_{tr}(\omega)]^{2} }{1 + [\omega_{c}^{z}
\tau_{tr}(\omega)]^{2} } B_{z} E_{x} . \nn \eqa Here, we performed an
expansion for the Berry curvature up to the second order since the
integration in the momentum space for the Berry curvature vanishes
when the integral expression contains an odd power for the Berry
curvature. Recall that the integral value is given by either $+1$
or $-1$, which depends on the chirality of the Weyl point
\cite{DTSon_Boltzmann}. $N_{F}$ is the density of states,
determined by the chemical potential.

The above expression becomes more simplified as follows \bqa &&
J_{x} = \mathcal{C} e^{2} N_{F} v_{F}^{2} \frac{ \tau_{tr}(T) }{1
+ [\omega_{c}^{z} \tau_{tr}(T)]^{2} } E_{x} + \mathcal{C}'
\frac{e^{4}}{c^{2}} N_{F} v_{F}^{2} \frac{ \tau_{tr}(T) }{1 +
[\omega_{c}^{z} \tau_{tr}(T)]^{2} } B_{z}^{2} E_{x} \nn && -
\mathcal{\tilde{C}} m \frac{e^{3}}{c} N_{F} v_{F}^{2}
\frac{[\omega_{c}^{z} \tau_{tr}(T)]^{2} \tau_{sc}(T)}{1 +
[\omega_{c}^{z} \tau_{tr}(T)]^{2} } B_{z} E_{x} ,  \eqa where
$\mathcal{C}$, $\mathcal{C}'$, and $\mathcal{\tilde{C}}$ are given
by integrals of momentum and frequency for the spectral function.
The first term is the conventional contribution for the
magnetoconductivity. On the other hand, other two terms result
from the Berry curvature.

In summary, the magnetoconductivity is \bqa && \sigma_{L}(B_{z},T)
= \mathcal{C} e^{2} N_{F} v_{F}^{2} \frac{ \tau_{tr}(T) }{1 +
[\omega_{c}^{z} \tau_{tr}(T)]^{2} } + \mathcal{C}'
\frac{e^{4}}{c^{2}} N_{F} v_{F}^{2} \frac{ \tau_{tr}(T) }{1 +
[\omega_{c}^{z} \tau_{tr}(T)]^{2} } B_{z}^{2} \nn && -
\mathcal{\tilde{C}} m \frac{e^{3}}{c} N_{F} v_{F}^{2}
\frac{[\omega_{c}^{z} \tau_{tr}(T)]^{2} \tau_{sc}(T)}{1 +
[\omega_{c}^{z} \tau_{tr}(T)]^{2} } B_{z} . \eqa

\subsection{$\boldsymbol{B} = B_{x} \boldsymbol{\hat{x}}$ and $\boldsymbol{E} = E_{x} \boldsymbol{\hat{x}}$}

Solving the quantum Boltzmann equation in the unconventional setup
of $\boldsymbol{B} = B_{x} \boldsymbol{\hat{x}}$ and
$\boldsymbol{E} = E_{x} \boldsymbol{\hat{x}}$, we find \bqa &&
\Lambda_{F}^{x}(\omega) = - e A(\boldsymbol{p}_{F},\omega)
\frac{\tau_{tr}(\omega)}{\tau_{sc}(\omega)} E_{x} , \eqa \bqa &&
\Lambda_{F}^{y}(\omega) = m e \frac{\omega_{c}^{x}
\tau_{tr}(\omega)}{\Bigl(1 + \frac{e}{c} B_{x}
\boldsymbol{\Omega}_{F}^{x}\Bigr)^{2} + [\omega_{c}^{x}
\tau_{tr}(\omega)]^{2}} \Bigl( - \boldsymbol{\Omega}_{F}^{z} +
\boldsymbol{\Omega}_{F}^{y} \frac{\omega_{c}^{x}
\tau_{tr}(\omega)}{1 + \frac{e}{c} B_{x}
\boldsymbol{\Omega}_{F}^{x}} \Bigr) E_{x} \nn && -
A(\boldsymbol{p}_{F},\omega) \frac{
\frac{\tau_{tr}(\omega)}{\tau_{sc}(\omega)}}{\Bigl(1 + \frac{e}{c}
B_{x} \boldsymbol{\Omega}_{F}^{x}\Bigr)^{2} + [\omega_{c}^{x}
\tau_{tr}(\omega)]^{2}} \Bigl\{ \frac{e^{2}}{c} + \frac{e^{2}}{c}
\Bigl( 1 + \frac{e}{c} B_{x} \boldsymbol{\Omega}_{F}^{x} \Bigr)
\Bigr\} \Bigl( \boldsymbol{\Omega}_{F}^{y} +
\boldsymbol{\Omega}_{F}^{z} \frac{\omega_{c}^{x}
\tau_{tr}(\omega)}{1 + \frac{e}{c} B_{x}
\boldsymbol{\Omega}_{F}^{x}} \Bigr) E_{x} B_{x} , \nn \eqa and \bqa &&
\Lambda_{F}^{z}(\omega) = m e \frac{\omega_{c}^{x}
\tau_{tr}(\omega)}{\Bigl(1 + \frac{e}{c} B_{x}
\boldsymbol{\Omega}_{F}^{x}\Bigr)^{2} + [\omega_{c}^{x}
\tau_{tr}(\omega)]^{2}} \Bigl( \boldsymbol{\Omega}_{F}^{y} +
\boldsymbol{\Omega}_{F}^{z} \frac{\omega_{c}^{x}
\tau_{tr}(\omega)}{1 + \frac{e}{c} B_{x}
\boldsymbol{\Omega}_{F}^{x}} \Bigr) E_{x} \nn && -
A(\boldsymbol{p}_{F},\omega) \frac{
\frac{\tau_{tr}(\omega)}{\tau_{sc}(\omega)}}{\Bigl(1 + \frac{e}{c}
B_{x} \boldsymbol{\Omega}_{F}^{x}\Bigr)^{2} + [\omega_{c}^{x}
\tau_{tr}(\omega)]^{2}} \Bigl\{ \frac{e^{2}}{c} + \frac{e^{2}}{c}
\Bigl( 1 + \frac{e}{c} B_{x} \boldsymbol{\Omega}_{F}^{x} \Bigr)
\Bigr\} \Bigl( \boldsymbol{\Omega}_{F}^{z} -
\boldsymbol{\Omega}_{F}^{y} \frac{\omega_{c}^{x}
\tau_{tr}(\omega)}{1 + \frac{e}{c} B_{x}
\boldsymbol{\Omega}_{F}^{x}} \Bigr) E_{x} B_{x} , \nn \eqa where
$\omega_{c}^{x} = \frac{e B_{x}}{m c}$ is the ``cyclotron"
frequency associated with the $B_{x}$ field. We notice that there
are $\boldsymbol{E}\cdot\boldsymbol{B} = E_{x} B_{x}$ terms, which
are topological in their origin.

Inserting these vertex distribution-functions into the current
formula, we obtain a rather complicated expression for the $x-$
component of the current, \bqa && J_{x} = e^{2} \frac{1}{\beta}
\sum_{i\omega} \int \frac{d^{3} \boldsymbol{p}}{(2\pi \hbar)^{3}}
\frac{1}{1 + \frac{e}{c} B_{x}\boldsymbol{\Omega}_{F}^{x}}
(v_{F}^{x})^{2} \Bigl( - \frac{\partial f(\omega)}{\partial
\omega} \Bigr) [A(\boldsymbol{p}_{F},\omega)]^{2}
\frac{\tau_{tr}(\omega)}{\tau_{sc}(\omega)} E_{x} \nn && +
\frac{e^{3}}{c} \frac{1}{\beta} \sum_{i\omega} \int \frac{d^{3}
\boldsymbol{p}}{(2\pi \hbar)^{3}} \frac{1}{1 + \frac{e}{c}
B_{x}\boldsymbol{\Omega}_{F}^{x}} (v_{\boldsymbol{p}}^{x})^{2}
\boldsymbol{\Omega}_{F}^{x} \Bigl( - \frac{\partial
f(\omega)}{\partial \omega} \Bigr)
[A(\boldsymbol{p}_{F},\omega)]^{2}
\frac{\tau_{tr}(\omega)}{\tau_{sc}(\omega)} B_{x} E_{x} \nn && -
\frac{m e^{3}}{c} \frac{1}{\beta} \sum_{i\omega} \int \frac{d^{3}
\boldsymbol{p}}{(2\pi \hbar)^{3}} \frac{1}{1 + \frac{e}{c} B_{x}
\boldsymbol{\Omega}_{F}^{x}} (v_{F}^{y})^{2} \Bigl( -
\frac{\partial f(\omega)}{\partial \omega} \Bigr)
A(\boldsymbol{p}_{F},\omega) \nn && \frac{\omega_{c}^{x}
\tau_{tr}(\omega)}{\Bigl(1 + \frac{e}{c} B_{x}
\boldsymbol{\Omega}_{F}^{x}\Bigr)^{2} + [\omega_{c}^{x}
\tau_{tr}(\omega)]^{2}} \Bigl( - \boldsymbol{\Omega}_{F}^{y}
\boldsymbol{\Omega}_{F}^{z} + (\boldsymbol{\Omega}_{F}^{y})^{2}
\frac{\omega_{c}^{x} \tau_{tr}(\omega)}{1 + \frac{e}{c} B_{x}
\boldsymbol{\Omega}_{F}^{x}} \Bigr) B_{x} E_{x} \nn && +
\frac{e^{4}}{c^{2}} \frac{1}{\beta} \sum_{i\omega} \int
\frac{d^{3} \boldsymbol{p}}{(2\pi \hbar)^{3}} \frac{2 +
\frac{e}{c} B_{x} \boldsymbol{\Omega}_{F}^{x}}{1 + \frac{e}{c}
B_{x}\boldsymbol{\Omega}_{F}^{x}} (v_{F}^{y})^{2} \Bigl( -
\frac{\partial f(\omega)}{\partial \omega} \Bigr)
[A(\boldsymbol{p}_{F},\omega)]^{2} \frac{
\frac{\tau_{tr}(\omega)}{\tau_{sc}(\omega)} \Bigl(
(\boldsymbol{\Omega}_{F}^{y})^{2} + \boldsymbol{\Omega}_{F}^{y}
\boldsymbol{\Omega}_{F}^{z} \frac{\omega_{c}^{x}
\tau_{tr}(\omega)}{1 + \frac{e}{c} B_{x}
\boldsymbol{\Omega}_{F}^{x}} \Bigr)}{\Bigl(1 + \frac{e}{c} B_{x}
\boldsymbol{\Omega}_{F}^{x}\Bigr)^{2} + [\omega_{c}^{x}
\tau_{tr}(\omega)]^{2}} E_{x} B_{x}^{2} \nn && - \frac{m e^{3}}{c}
\frac{1}{\beta} \sum_{i\omega} \int \frac{d^{3}
\boldsymbol{p}}{(2\pi \hbar)^{3}} \frac{1}{1 + \frac{e}{c} B_{x}
\boldsymbol{\Omega}_{F}^{x}} (v_{F}^{z})^{2} \Bigl( -
\frac{\partial f(\omega)}{\partial \omega} \Bigr)
A(\boldsymbol{p}_{F},\omega) \nn && \frac{\omega_{c}^{x}
\tau_{tr}(\omega)}{\Bigl(1 + \frac{e}{c} B_{x}
\boldsymbol{\Omega}_{F}^{x}\Bigr)^{2} + [\omega_{c}^{x}
\tau_{tr}(\omega)]^{2}} \Bigl( \boldsymbol{\Omega}_{F}^{y}
\boldsymbol{\Omega}_{F}^{z} + (\boldsymbol{\Omega}_{F}^{z})^{2}
\frac{\omega_{c}^{x} \tau_{tr}(\omega)}{1 + \frac{e}{c} B_{x}
\boldsymbol{\Omega}_{F}^{x}} \Bigr) B_{x} E_{x} \nn && +
\frac{e^{4}}{c^{2}} \frac{1}{\beta} \sum_{i\omega} \int
\frac{d^{3} \boldsymbol{p}}{(2\pi \hbar)^{3}} \frac{2 +
\frac{e}{c} B_{x} \boldsymbol{\Omega}_{F}^{x}}{1 + \frac{e}{c}
B_{x} \boldsymbol{\Omega}_{F}^{x}} (v_{F}^{z})^{2} \Bigl( -
\frac{\partial f(\omega)}{\partial \omega} \Bigr)
[A(\boldsymbol{p}_{F},\omega)]^{2} \frac{
\frac{\tau_{tr}(\omega)}{\tau_{sc}(\omega)} \Bigl(
(\boldsymbol{\Omega}_{F}^{z})^{2} - \boldsymbol{\Omega}_{F}^{y}
\boldsymbol{\Omega}_{F}^{z} \frac{\omega_{c}^{x}
\tau_{tr}(\omega)}{1 + \frac{e}{c} B_{x}
\boldsymbol{\Omega}_{F}^{x}} \Bigr)}{\Bigl(1 + \frac{e}{c} B_{x}
\boldsymbol{\Omega}_{F}^{x}\Bigr)^{2} + [\omega_{c}^{x}
\tau_{tr}(\omega)]^{2}} E_{x} B_{x}^{2} . \nn \eqa

Expanding the above expression up to the second order for the
Berry curvature, we obtain \bqa && J_{x} \approx e^{2}
\frac{1}{\beta} \sum_{i\omega} \int \frac{d^{3}
\boldsymbol{p}}{(2\pi \hbar)^{3}} (v_{F}^{x})^{2} \Bigl( -
\frac{\partial f(\omega)}{\partial \omega} \Bigr)
[A(\boldsymbol{p}_{F},\omega)]^{2}
\frac{\tau_{tr}(\omega)}{\tau_{sc}(\omega)} E_{x} \nn && - 2
\frac{m e^{3}}{c} \frac{1}{\beta} \sum_{i\omega} \int \frac{d^{3}
\boldsymbol{p}}{(2\pi \hbar)^{3}} (v_{F}^{y})^{2} \Bigl( -
\frac{\partial f(\omega)}{\partial \omega} \Bigr)
A(\boldsymbol{p}_{F},\omega) \frac{
(\boldsymbol{\Omega}_{F}^{y})^{2} \omega_{c}^{x}
\tau_{tr}(\omega)}{1 + [\omega_{c}^{x} \tau_{tr}(\omega)]^{2}}
[\omega_{c}^{x} \tau_{tr}(\omega)] B_{x} E_{x} \nn && + 2
\frac{e^{4}}{c^{2}} \frac{1}{\beta} \sum_{i\omega} \int
\frac{d^{3} \boldsymbol{p}}{(2\pi \hbar)^{3}} (v_{F}^{y})^{2}
\Bigl( - \frac{\partial f(\omega)}{\partial \omega} \Bigr)
[A(\boldsymbol{p}_{F},\omega)]^{2} \frac{
\frac{\tau_{tr}(\omega)}{\tau_{sc}(\omega)}
(\boldsymbol{\Omega}_{F}^{y})^{2} }{1 + [\omega_{c}^{x}
\tau_{tr}(\omega)]^{2}} E_{x} B_{x}^{2} \nn && = \mathcal{C} N_{F}
e^{2} v_{F}^{2} \tau_{tr}(T) E_{x} + 2 \mathcal{C}'
\frac{e^{4}}{c^{2}} N_{F} v_{F}^{2} \frac{ \tau_{tr}(T)}{1 +
[\omega_{c}^{x} \tau_{tr}(T)]^{2}} B_{x}^{2} E_{x} - 2
\mathcal{C}'' \frac{m e^{3}}{c} N_{F} v_{F}^{2} \frac{\tau_{sc}(T)
[\omega_{c}^{x} \tau_{tr}(T)]^{2}}{1 + [\omega_{c}^{x}
\tau_{tr}(T)]^{2}} B_{x} E_{x} . \nn \eqa The first term is also the
conventional contribution near the Fermi surface, but there is no
dependence for magnetic fields. This is certainly expected because
the magnetic field is in the same direction as the electric field.
On the other hand, the second contribution originates from the
topological $\boldsymbol{E}\cdot\boldsymbol{B}$ term. The third
term is also anomalous, which results from the Berry curvature but
not from the $\boldsymbol{E}\cdot\boldsymbol{B}$ term.

In summary, the ``longitudinal" magnetoconductivity is \bqa &&
\sigma_{L}(B_{x},T) = \mathcal{C} N_{F} e^{2} v_{F}^{2}
\tau_{tr}(T) + 2 \mathcal{C}' \frac{e^{4}}{c^{2}} N_{F} v_{F}^{2}
\frac{ \tau_{tr}(T)}{1 + [\omega_{c}^{x} \tau_{tr}(T)]^{2}}
B_{x}^{2} - 2 \mathcal{C}'' \frac{m e^{3}}{c} N_{F} v_{F}^{2}
\frac{\tau_{sc}(T) [\omega_{c}^{x} \tau_{tr}(T)]^{2}}{1 +
[\omega_{c}^{x} \tau_{tr}(T)]^{2}} B_{x} . \nn \eqa

\subsection{Discussion}

In order to compare the expression of the ``longitudinal"
magnetoconductivity with that used in our fitting for the
experimental data, we rewrite the above expression as follows \cite{Jho_Kim} \bqa
&& \sigma_{L}(B_{x},T) = (1 + \mathcal{C}_{W} B_{x}^{2})
\sigma_{n}(T) , \eqa where $\sigma_{n}(T) = \mathcal{C} N_{F}
e^{2} v_{F}^{2} \tau_{tr}(T)$ is the normal conductivity and
$\mathcal{C}_{W} = 2 (\mathcal{C}' / \mathcal{C}) (e^{2}/c^{2})$
is a positive constant. Fitting to our experimental data, we
replace this normal contribution with the conductivity that
contains weak-antilocalization corrections, named as
$\sigma_{WAL}(B_{x},T)$. In addition, we introduce another normal
conductivity, denoted by $\sigma_{n}(T)$, which originates from
other bands at the chemical potential. Although we use the same
symbol as the above, do not confuse the apparent difference. As a
result, we propose the ``longitudinal" magnetoconductivity in the
presence of the weak-antilocalization correction as follows \bqa
&& \sigma_{L}(B_{x},T) = (1 + \mathcal{C}_{W} B_{x}^{2})
\sigma_{WAL}(B_{x},T) + \sigma_{n}(T) , \eqa where the overall
factor $\mathcal{C}_{W} B_{x}^{2}$ is purely topological in
its origin. Actually, what we have done in this supplementary
material is to derive this overall factor. In the conventional
setup with $\boldsymbol{B} = B_{z} \boldsymbol{\hat{z}}$ and
$\boldsymbol{E} = E_{x} \boldsymbol{\hat{x}}$, we suggest
$\sigma_{L}(B_{z},T) = \sigma_{WAL}(B_{z},T) +
\sigma_{n}(B_{z},T)$, where $\sigma_{n}(B_{z},T)$ results from
other bands. This expression turns out to reproduce the
experimental data even quantitatively well.

\section{Physical interpretation for anisotropic transport coefficients}

Although anisotropic transport coefficients in the longitudinal configuration
($\boldsymbol{B} = B_{x} \boldsymbol{\hat{x}}$ and $\boldsymbol{E} = E_{x} \boldsymbol{\hat{x}}$) have been investigated based on the quantum Boltzmann equation approach with the introduction of the Adler-Bell-Jackiw anomaly via the semi-classical equations of motion, technical complexity of this methodology does not allow us to have an intuitive physical picture for anomalous behaviors in such transport coefficients. In this respect it is necessary to confirm physics of the Adler-Bell-Jackiw anomaly based on a simpler framework. In this subsection we rederive such anomalous transport coefficients, resorting to the semi-classical equations of motion only. This approach is basically the same as that for the Hall effect (via the Lorentz force) in the elemetary solid-state physics, except for the introduction of the topological $\theta-$ term and the contribution from the Berry curvature.

We start from the solution of the semi-classical equations of motion
\bqa && \boldsymbol{\dot{p}} = \Bigl( 1 + \frac{e}{c} \boldsymbol{B} \cdot
\boldsymbol{\Omega}_{\boldsymbol{p}} \Bigr)^{-1} \Bigl\{ e
\boldsymbol{E} + \frac{e}{m c} \boldsymbol{p}
\times \boldsymbol{B} + \frac{e^{2}}{c} (\boldsymbol{E} \cdot
\boldsymbol{B}) \boldsymbol{\Omega}_{\boldsymbol{p}} \Bigr\} = - \frac{\boldsymbol{p}}{\tau} ,
\eqa where the momentum of an electron wave-packet is relaxed after the mean-free time $\tau$.
Here, $\tau = \tau_{intra} \tau_{inter} / (\tau_{intra} + \tau_{inter}) \sim \tau_{intra}$
in the semi-classical regime of $\tau_{intra} \ll \tau_{inter}$, where $\tau_{intra(inter)}$
is the mean-free time associated with intra- (inter-) node scattering due to impurities \cite{DTSon_Boltzmann}.

Solving this equation of motion to obtain the momentum as a function of both electric and magnetic fields, we find the corresponding electrical current, given by
\bqa && \boldsymbol{J} = n e \boldsymbol{\dot{r}} = n e \Bigl( 1 + \frac{e}{c}
\boldsymbol{B} \cdot \boldsymbol{\Omega}_{\boldsymbol{p}}
\Bigr)^{-1} \Bigl\{ \frac{\boldsymbol{p}}{m} + e
\boldsymbol{E} \times \boldsymbol{\Omega}_{\boldsymbol{p}} +
\frac{e}{m c} (\boldsymbol{\Omega}_{\boldsymbol{p}} \cdot
\boldsymbol{p}) \boldsymbol{B} \Bigr\} , \eqa where the other solution for the semi-classical equations of motion is utilized. It is straightforward to obtain transport coefficients from this expression. It is essential to notice that there are two more contributions for electric currents than the conventional momentum-proportional term, which turn out to be nonzero only when the Berry curvature exists. In particular, the last term proportional to the applied magnetic field gives rise to anomalous contributions for the unconventional Hall coefficients, here the longitudinal Hall resistivity, where the topological $\boldsymbol{E} \cdot
\boldsymbol{B}$ term in the solution of the momentum plays an important role.


When electric fields are applied in parallel with magnetic fields, saying the $x-$direction, the semiclassical equation of motion
for the momentum is expressed as follows by each component,
\bqa && - \frac{p_{x}}{\tau} = \Bigl( 1 + \frac{e}{c} B_{x} {\Omega}_{\boldsymbol{p}}^{x} \Bigr)^{-1} \Bigl\{ e
E_{x} + \frac{e^{2}}{c} (E_{x} B_{x}) {\Omega}_{\boldsymbol{p}}^{x} \Bigr\} , \\ &&
- \frac{p_{y}}{\tau} = \Bigl( 1 + \frac{e}{c} B_{x} {\Omega}_{\boldsymbol{p}}^{x} \Bigr)^{-1}
\Bigl\{ e E_{y} + \omega_{c}^{x} p_{z} + \frac{e^{2}}{c} (E_{x} B_{x}) {\Omega}_{\boldsymbol{p}}^{y} \Bigr\} , \\ &&
- \frac{p_{z}}{\tau} = \Bigl( 1 + \frac{e}{c} B_{x} {\Omega}_{\boldsymbol{p}}^{x} \Bigr)^{-1}
\Bigl\{ - \omega_{c}^{x} p_{y} + \frac{e^{2}}{c} (E_{x} B_{x}) {\Omega}_{\boldsymbol{p}}^{z} \Bigr\} , \eqa where
$\omega_{c}^{x} = \frac{e B_{x}}{m c}$ is the cyclotron frequency for the $x-$directional magnetic field.

In order to obtain the Hall coefficient, we introduce the following condition \bqa && J_{y} = n e \Bigl( 1 + \frac{e}{c}
B_{x} {\Omega}_{\boldsymbol{p}}^{x} \Bigr)^{-1} \Bigl( \frac{p_{y}}{m} - e
E_{x} {\Omega}_{\boldsymbol{p}}^{z} \Bigr) \longrightarrow 0 \nonumber \eqa from the definition of the current, resulting in
\bqa && p_{y} = e m {\Omega}_{\boldsymbol{p}}^{z} E_{x} . \eqa

Two coupled equations for $p_{y}$ and $p_{z}$ give the information on the Hall voltage with the use of the condition for the Hall coefficient,
%
%
%
\bqa && E_{y} = - \frac{m}{\tau} {\Omega}_{\boldsymbol{p}}^{z} \Bigl( 1 + \frac{e}{c} B_{x} {\Omega}_{\boldsymbol{p}}^{x} \Bigr) \Bigl\{1 + \Bigl( 1 + \frac{e}{c} B_{x} {\Omega}_{\boldsymbol{p}}^{x} \Bigr)^{-2} (\omega_{c}^{x} \tau)^{2} \Bigr\} E_{x} \nn && + \frac{e}{c} \Bigl\{ (\omega_{c}^{x} \tau)
\Bigl( 1 + \frac{e}{c} B_{x} {\Omega}_{\boldsymbol{p}}^{x} \Bigr)^{-1} {\Omega}_{\boldsymbol{p}}^{z}
- {\Omega}_{\boldsymbol{p}}^{y} \Bigr\} (E_{x} B_{x} ) . \eqa

Considering the definition of the Hall coefficient, $E_{y} = \rho_{yx}(B_{x}) J_{x}$, we introduce an electric current of the $x-$direction,
\bqa && J_{x} = n e \Bigl( 1 + \frac{e}{c} B_{x} {\Omega}_{\boldsymbol{p}}^{x} \Bigr)^{-1} \Bigl\{ \frac{p_{x}}{m} +
\frac{e}{m c} (\boldsymbol{\Omega}_{\boldsymbol{p}} \cdot \boldsymbol{p}) B_{x} \Bigr\} . \eqa

Solving the equations of motion, we obtain
\bqa && p_{x} = - e \tau E_{x} , \\ &&
p_{z} = e \tau \Bigl( 1 + \frac{e}{c} B_{x} {\Omega}_{\boldsymbol{p}}^{x} \Bigr)^{-1}
\Bigl(\frac{e}{c} {\Omega}_{\boldsymbol{p}}^{z} E_{x} - \frac{e}{c} (E_{x} B_{x}) {\Omega}_{\boldsymbol{p}}^{z} \Bigr) . \eqa

Inserting these solutions with the condition $E_{y}(E_{x},B_{x})$ for the Hall coefficient into the expression of the $x-$directional current, we find
%
%
the longitudinal magnetoconductivity
\bqa && \sigma_{L}(B_{x}) = \Bigl\{ 1 + \frac{e^{2}}{c^{2}} \frac{{\Omega}_{\boldsymbol{p}}^{z 2}}{\Bigl( 1 + \frac{e}{c}
B_{x} {\Omega}_{\boldsymbol{p}}^{x} \Bigr)^{2}} B_{x}^{2} \Bigr\} \sigma - n e^{2} \frac{e}{c} \frac{{\Omega}_{\boldsymbol{p}}^{z} {\Omega}_{\boldsymbol{p}}^{y}}{1 + \frac{e}{c}
B_{x} {\Omega}_{\boldsymbol{p}}^{x}} B_{x} - \sigma \frac{e^{2}}{c^{2}} \frac{{\Omega}_{\boldsymbol{p}}^{z 2}}{\Bigl( 1 + \frac{e}{c}
B_{x} {\Omega}_{\boldsymbol{p}}^{x} \Bigr)^{2}  } B_{x} , \nn \eqa where $\sigma = \frac{n e^{2} \tau}{m}$ is the dc Drude conductivity.
Experimentally, we symmetrize this transport coefficient and obtain
\bqa && \sigma_{exp}^{L} (B_{x}) = \frac{\sigma_{L}(B_{x}) + \sigma_{L}(- B_{x})}{2}
\approx \Bigl( 1 + \frac{e^{2}}{c^{2}} B_{x}^{2} \Bigr) \sigma , \eqa
essentially the same as the expression from the quantum Boltzmann equation approach. It is easy to observe that the $B_x^2$ term results from the $\boldsymbol{E} \cdot \boldsymbol{B}$ term from this equation of motion approach. In spite of this consistency, we would like to point out that the coefficient $\mathcal{C}_{W}$ in Eq. (29) does not appear in this much simplified treatment.

\end{document}